\documentclass[sigconf,9pt]{acmart}

\usepackage[english]{babel}
\usepackage{blindtext}
\usepackage{array}
\usepackage{enumitem}
\usepackage{subfigure}
\usepackage{makecell}
\usepackage{multirow}
\usepackage{multirow}
\usepackage{caption}
\usepackage{threeparttable}
\setcopyright{acmlicensed}
\copyrightyear{2024}
\acmYear{2024}
\acmDOI{https://doi.org/10.1145/3651890.3672268}
\acmYear{2024}\copyrightyear{2024}
\acmConference[ACM SIGCOMM '24]{ACM SIGCOMM 2024 Conference}{August 4--8, 2024}{Sydney, NSW, Australia}
\acmBooktitle{ACM SIGCOMM 2024 Conference (ACM SIGCOMM '24), August 4--8, 2024, Sydney, NSW, Australia}
\acmDOI{10.1145/3651890.3672268}
\acmISBN{979-8-4007-0614-1/24/08}

\newcommand{\Revision}[1]{\textcolor{black}{#1}}
\newcommand{\SecRevision}[1]{\textcolor{black}{#1}}
\acmISBN{979-8-4007-0614-1/24/08}

\acmPrice{}

\begin{document}
\title[\texttt{NetLLM}:  Adapting Large Language Models for Networking]{\texttt{NetLLM}: Adapting Large Language Models for Networking}

\author{Duo Wu$^1$, Xianda Wang$^1$, Yaqi Qiao$^1$, Zhi Wang$^2$, Junchen Jiang$^3$, Shuguang Cui$^1$, Fangxin Wang$^1$}\authornote{Fangxin Wang is the corresponding author: wangfangxin@cuhk.edu.cn.}
\def \authors{Duo Wu, Xianda Wang, Yaqi Qiao, Zhi Wang, Junchen Jiang, Shuguang Cui, Fangxin Wang}
\affiliation{%
\institution{$^1$SSE and FNii, The Chinese University of Hong Kong, Shenzhen\\$^2$SIGS, Tsinghua University \quad $^3$The University of Chicago}
}

\renewcommand{\shortauthors}{Wu et al.}

\begin{abstract}
    
    Many networking tasks now employ deep learning (DL) to solve complex prediction and optimization problems. However, current design philosophy of DL-based algorithms entails intensive engineering overhead due to the manual design of deep neural networks (DNNs) for different networking tasks. Besides, DNNs tend to achieve poor generalization performance on unseen data distributions/environments.

    Motivated by the recent success of large language models (LLMs), this work studies the LLM adaptation for networking to explore a more sustainable design philosophy. With the powerful pre-trained knowledge, the LLM is promising to serve as the foundation model to achieve ``one model for all tasks'' with even better performance and stronger generalization. In pursuit of this vision, we present \texttt{NetLLM}, the first framework that provides a coherent design to harness the powerful capabilities of LLMs with low efforts to solve networking problems. Specifically, \texttt{NetLLM} empowers the LLM to effectively process multimodal data in networking and efficiently generate task-specific answers. Besides, \texttt{NetLLM} drastically reduces the costs of fine-tuning the LLM to acquire domain knowledge for networking. Across three networking-related use cases - viewport prediction, adaptive bitrate streaming and cluster job scheduling, we  showcase that the \texttt{NetLLM}-adapted LLM significantly outperforms state-of-the-art algorithms.
    
\end{abstract}

\ccsdesc[500]{Networks ~ Application layer protocols}
\ccsdesc[500]{Computing methodologies ~ Supervised learning}
\ccsdesc[500]{Computing methodologies ~ Reinforcement learning}

\keywords{Deep Learning, Network Optimization, Video Streaming, Job Scheduling, Large Language Model Adaptation}

\maketitle

 \captionsetup[figure]{labelfont={bf, small},textfont={bf, small}}%
 \captionsetup[table]{labelfont={bf, small},textfont={bf, small}}%

\section{Introduction}

\subsection{The Main Roadmap so far}
Over the past decades, \textit{rule-based algorithms} built on handcrafted control rules have played an important role in optimizing network systems~\cite{arun2018copa,robert2014multi,bai2015information,yin2015control}.
For instance, Copa~\cite{arun2018copa} adjusts sending rates for congestion control based on measured queueing delay, while PANDA~\cite{li2014probe} switches video streaming bitrates based on heuristically estimated bandwidth.
However, these algorithms heavily rely on \textit{rule engineering}, which involves intensive human efforts to devise, implement and validate the control rules for network optimization~\cite{li2018auto,meng2020interpreting,mao2019learning,mao2017neural}.

In recent years, the advancements of deep learning have prompted extensive research into \textit{learning-based algorithms} for networking. These algorithms design and train deep neural networks (DNNs) with supervised learning (SL)~\cite{wang2020supervised} or reinforcement learning (RL)~\cite{sutton2018reinforcement} techniques to automatically discover networking solutions, thus eliminating the need of rule engineering.
Specifically, SL is widely adopted to train DNNs for prediction tasks in networking, such as traffic classification~\cite{lin2022bert,pacheco2018towards} and bandwidth prediction~\cite{abdelhak2019bandwidth,mei2020realtime}. On the flip side, RL is commonly employed to solve decision-making problems in networking, including congestion control~\cite{abbasloo2020classic,yen2023computers}, adaptive bitrate streaming (ABR)~\cite{mao2017neural,kan2022improving} and cloud cluster job scheduling (CJS)~\cite{mao2019learning,pen2021dl2}. Thanks to the strong capability of DNNs in function approximation, learning-based algorithms have demonstrated significant improvement over handcrafted rule-based algorithms.

Despite their promising potential, existing learning-based algorithms still suffer from two key limitations:
\begin{itemize}[leftmargin=0.8em]
    
    \item \textbf{High model engineering costs.} \SecRevision{It has been shown that the design of DNN architecture is crucial to the final performance~\cite{evolving2024robert,comprehensive2021ren}.}
    Therefore, the focus of learning-based algorithms has shifted from rule engineering to \textit{model engineering}. Their success is heavily dependent on engineering DNN models for the target networking tasks, which, however, can be difficult and labor-intensive due to the complex structures of DNNs~\cite{meng2020interpreting}. To make things worse, the diversity of networking tasks also prevents sharing the same DNN model across different tasks. 
    This necessitates designing specialized DNNs for different tasks (i.e., \textit{one model for one task}), thus further increasing the engineering costs. 
    \SecRevision{Although some recent works attempt to introduce structured Transformer~\cite{vaswani2017attention} into model design, they still necessitate manual tuning of the Transformer architecture (e.g., the number of attention blocks and attention heads)~\cite{wu2023mansy}, or even the design of specialized tokenization scheme~\cite{guthula2023netfound,meng2023netgpt} and attention mechanism~\cite{liu2023cav3,guthula2023netfound}, thus leading to high engineering costs. }
    
    \item \textbf{Low generalization.} DNNs trained on specific data distributions/environments may struggle to perform well or even worse than conventional rule-based algorithms on unseen data distributions/environments~\cite{yan2020learning}. 
    For example, an ABR model trained on smooth network conditions often achieves poor performance on network environments with dynamic bandwidth fluctuations~\cite{kan2022improving}.
    The lack of generalization can ultimately hinder the widespread deployment of learning-based algorithms in practice~\cite{xia2022genet,yen2023computers}, as network operators will suspect their superiority over the incumbent rule-based algorithms in production environments.
\end{itemize}
\vspace{-0.2cm}

\subsection{New Opportunities and Challenges}
Utilizing a single generic model for different tasks has been recognized as a significant approach to mitigate the costs of handcrafting specialized DNNs for each task and enhance generalization~\cite{reed2022a}. 
This is exemplified by the recent popular large language models (LLMs) such as ChatGPT~\cite{chatgpt}, Falcon~\cite{penedo2023refinedweb} and  Llama2~\cite{touvron2023llama} in the field of natural language processing (NLP). 
With billions of parameters pre-trained on massive data to absorb extensive knowledge, LLMs have demonstrated extraordinary capacity in conversations, reasoning and text generation in NLP~\cite{wei2022finetuned}. 
\Revision{What's more, they also exhibit emergent abilities that were not explicitly programmed into them during pre-training, such as planning, pattern mining, problem solving and generalization to unseen conditions~\cite{xi2023rise,rt22023zitkovich}. These abilities have been proven to be transferable to other domains, including robotics~\cite{driess2023palme}, chip design~\cite{liu2023chipnemo}, and protein structure prediction~\cite{lin2023evolutionary}. For instance, researchers have shown that LLMs can generate goal-oriented plans for robotic control, adjusting plans in response to environment changes, and generalize to previously unseen operating environments~\cite{rt22023zitkovich}.}

\Revision{Motivated by these inspiring outcomes, we believe that LLMs can serve as foundation models for networking, as many networking tasks can also benefit from their emergent abilities. In the context of ABR, for example, the planning ability of LLMs can be utilized for better bitrate decisions to optimize the video streaming sessions based on the changing network conditions. Furthermore, their generalization capability can be harnessed to generalize across diverse network environments.}
Therefore, we envision LLM as the key to achieving \textit{one model for all tasks}, with little handcraft costs and strong generalization. We try to answer the following key question: \textit{can we embrace the era of LLMs and adapt LLMs to solve various networking tasks efficiently and effectively?}

Unfortunately, as revealed by our analysis in $\S$\ref{sec:motivation}, the adaptation of LLMs for networking faces the following challenges.


\begin{itemize}[leftmargin=0.8em]
     \item \textbf{Large input modality gap.} In networking tasks, various information observed from the system is collected as inputs for networking algorithms. 
     However, the modalities of these inputs differ significantly from plain text, i.e., the native input modality supported by  LLMs\footnote{Although multimodal LLMs have emerged recently (e.g., GPT4~\cite{openai2023gpt4}), their supported input modalities  are still limited (mainly vision or audio), hindering directly applying them to process networking task inputs.}~\cite{touvron2023llama,zhang2022opt}. \Revision{For example, in ABR, network throughputs and delay are often collected for bitrate decision~\cite{mao2017neural}, which exhibit unique time-varying data patterns typically not found in natural text.} This discrepancy prevents LLMs from effectively processing the input information of networking tasks.

    \item \textbf{Inefficiency of answer generation.} LLMs generate answers using a language modeling (LM) head to predict words (tokens) \textit{one by one} (see Figure~\ref{fig:llm_mechnism})~\cite{kwon2023efficient}. While this approach is well-suited in NLP, it presents several drawbacks in the networking domain. First, LLMs are prone to \textit{hallucination} due to the inherent uncertainty of token prediction~\cite{ji2023survey,li2023halueval}. Their generated answers for networking tasks may seem correct but physically invalid (e.g., a nonexistent bitrate for video download in ABR), which can eventually impair the reliability of network systems. Second, since tokens are predicted one at a time, LLMs often require multiple inferences to generate a complete answer, thus incurring high answer generation latency (i.e., the time to generate a complete answer). Consequently, LLMs may fail to quickly generate answers to respond to the system changes (e.g., switching to a lower bitrate when network bandwidth becomes scarce). 
    
    \item \textbf{High adaptation costs.} The large domain gap between networking and NLP necessitates fine-tuning LLMs to acquire domain-specific knowledge for effective adaptation. 
    However, the adaptation costs can be prohibitively high, especially when fine-tuning LLMs for decision-making tasks (e.g., ABR~\cite{kan2022improving} and CJS~\cite{mao2019learning}) where RL is employed to solve the system optimization problems. Specifically, RL-based decision-making tasks require the active interaction between LLMs and environments (e.g., network environments with varying bandwidth or workload patterns) to collect experiences for performance optimization~\cite{mao2019learning,xia2022genet}. Due to the large parameter size of LLMs, the interaction process can be excessively time-consuming, introducing a large amount of additional training time. 
    What's worse, the adaptation costs can further increase if fine-tuning the full parameters of LLMs, which is known to be resource-intensive~\cite{lin2023evolutionary,liu2023chipnemo}.
\end{itemize}

\subsection{Design and Contributions}


\SecRevision{To overcome the aforementioned challenges, this work proposes \texttt{NetLLM}, \textit{the first framework that efficiently adapts LLMs for networking}. \texttt{NetLLM} stands out as the first to take LLMs the extra mile in context of networking and provides a coherent design to utilize their powerful capabilities to generalize to various tasks with low efforts. It enables the efficient utilization of a single LLM (e.g., Llama2~\cite{touvron2023llama}) as the foundation model without any modifications to tackle a wide range of networking tasks and meanwhile achieves enhanced performance. To accomplish this, \texttt{NetLLM} designs a multimodal encoder to enable the LLM to process multimodal data in networking and implements a networking head to improve the efficiency of answer generation. Furthermore, while fine-tuning the LLM and these two modules on the target task to acquire domain knowledge is necessary, \texttt{NetLLM} designs an efficient data-driven low-rank networking adaptation (DD-LRNA) scheme to drastically reduce the fine-tuning costs. Specifically, \texttt{NetLLM} incorporates the following three core design modules to efficiently adapt LLMs for networking.}

\begin{itemize}[leftmargin=0.8em]
    \item \textbf{Multimodal encoder.} 
    \texttt{NetLLM} designs an efficient multimodal encoder at the input side of the LLM to effectively process the multimodal input information of networking tasks. 
    \Revision{The goal of this module is to automatically project task inputs to the same feature space as language tokens, so that the LLM can understand and utilize these inputs for task solving. To achieve this, it first uses modality-specific feature encoders to extract features from raw inputs of various modalities involved in networking.} It then leverages trainable layers to project these features into token-like embedding vectors, which can be directly fed into the LLM for effective processing. 
    \item \textbf{Networking head.} 
    To enable efficient answer generation, \texttt{NetLLM} removes the default LM head used by the LLM for token prediction. Instead, it introduces various networking heads at the output side of the LLM to generate answers for specific networking tasks.
    Each networking head is essentially a lightweight trainable projector that  maps the output features of the LLM directly into task-specific answers. 
    \Revision{In other words, they eliminate token prediction and allow direct answer generation from the valid range of possible answers (e.g., selecting a bitrate from candidate options in ABR).} This enables the LLM to generate a valid answer in a single inference, thus ensuring its reliability for networking and significantly reducing generation latency.
    
    \item  \textbf{DD-LRNA.}
    To reduce the fine-tuning costs, \texttt{NetLLM} designs a DD-LRNA scheme for the LLM to efficiently acquire domain knowledge for networking. Specifically, DD-LRNA incorporates a data-driven adaptation pipeline to adapt the LLM for both prediction and  decision-making tasks.
    In particular, for decision-making tasks, it employs the efficient data-driven RL technique~\cite{yen2023computers,agarwal2020optimistic,prudencio2023survey} to eliminate the time-consuming interaction between the LLM and environments. 
    \Revision{It collects an experience pool as training dataset with existing networking algorithms and fine-tunes the LLM over such dataset in the data-driven manner.}  
    Besides, inspired by the advanced parameter-efficient fine-tune  technique~\cite{ding2023parameter,fu2023on}, DD-LRNA introduces a set of additional  trainable low-rank matrices for the LLM to learn networking knowledge. Since the low-rank matrices only account for 0.31\% of the total parameters, the fine-tune costs are greatly reduced, with the reduction of 60.9\% GPU memory and 15.1\% training time. 
    
\end{itemize}

\texttt{NetLLM} has the following important properties. i) \textit{Compatibility}: It is a generic framework that can be applied to adapt different LLMs for various networking tasks. 
ii) \textit{Reliability}: It addresses the hallucination issue and ensures the LLM generated answers to be always valid. iii) \textit{Efficiency}: The DD-LRNA scheme significantly reduces the costs of fine-tuning LLM to learn domain knowledge and the networking heads also reduce the answer generation latency. \SecRevision{iv) \textit{Transferability}: \texttt{NetLLM} offers an effective solution for utilizing LLMs to solve prediction and decision-making tasks at low costs. These two types of tasks, in fact, span across diverse domains such as medicine~\cite{clusmann2023future}, finance~\cite{li2023large}, and telecommunication~\cite{zhou2024large}. Hence, despite its original focus on networking, \texttt{NetLLM} holds the potential to be applied and transferred to other fields.} 

We have implemented \texttt{NetLLM}\footnote{The codes available at: \url{https://github.com/duowuyms/NetLLM}.} for three networking tasks: viewport prediction (VP)~\cite{rondon2021track} for immersive video streaming, adaptive bitrate streaming (ABR)~\cite{xia2022genet}, and cluster job scheduling (CJS)~\cite{mao2019learning}. We believe these representative tasks cover the main input modalities of networking problems and span from prediction tasks to decision-making tasks ($\S$\ref{sec:motivation}). 
Through extensive trace-driven simulation and real-world tests, we demonstrate the effectiveness of \texttt{NetLLM} in LLM adaptation for networking. We showcase that across the three use cases, the \texttt{NetLLM}-adapted LLM  significantly outperforms state of the arts with performance improvements of 10.1-36.6\% for VP,  14.5-36.6\% for ABR, 6.8-41.3\% for CJS, in terms of their respective performance metrics. Besides, our empirical test results also show that we can efficiently utilize the extensive knowledge of the LLM with our proposed \texttt{NetLLM} framework to achieve stronger generalization on unseen testing environments.

The contributions of this paper are summarized as follows:
\begin{itemize}[leftmargin=0.95em]
    \item We identify the key challenges of adapting LLMs for networking and demonstrate that some natural alternatives fall short in addressing these challenges ($\S$\ref{sec:motivation}). 
    \item We then design \texttt{NetLLM}, the first LLM adaptation framework for networking that incorporates a multimodal encoder module to encode multimodal task inputs, networking head module to directly generate answers and a DD-LRNA scheme to  reduce the adaptation costs ($\S$\ref{sec:netllm_design}).
    \item We extensively evaluate \texttt{NetLLM} across three networking tasks. We showcase that the LLM adapted by our framework can significantly surpass state-of-the-art algorithms and achieve superior generalization performance. 
    We also conduct a deep dive into \texttt{NetLLM} to provide an in-depth understanding of it ($\S$\ref{sec:evaluation}).
\end{itemize}

\section{Background}
\subsection{Learning-Based Networking Algorithms }
\label{subsec:networking_bg} 

Learning-based algorithms design and train deep neural networks (DNNs) to efficiently learn to solve networking tasks~\cite{mei2020realtime,xia2022genet,abbasloo2021wanna}. In particular, there are two learning paradigms commonly adopted to enable the learning process: supervised learning (SL)~\cite{wang2020supervised} and reinforcement learning (RL)~\cite{sutton2018reinforcement}. SL is widely employed for prediction tasks in networking, such as traffic classification~\cite{lin2022bert,pacheco2018towards}, bandwidth prediction~\cite{yan2020learning,mei2020realtime} and viewport prediction~\cite{guimard2022deep,rondon2021track}. It trains DNNs with specific datasets to optimize a pre-defined loss function, so that once trained, DNNs can be used for efficient prediction to assist the system control. For example, Yan et al.~\cite{yan2020learning} train a DNN model over real-world bandwidth datasets to predict the future bandwidth for bitrate control on video clients. On the flip side, RL is well-suited for sequential decision-making tasks in networking, including congestion control~\cite{abbasloo2020classic,yen2023computers}, video streaming~\cite{mao2017neural,kan2022improving} and cluster job scheduling~\cite{mao2019learning,pen2021dl2}. In these tasks, DNNs actively interact with the environments to collect experiences, then use them to optimize task-specific reward functions so that the performance of network systems can be optimized. For instance, Mao et al.~\cite{mao2019learning} employ RL to train a DNN model to allocate resources to job requests so as to maximize the utilization of computing resources in a distributed computing cluster.

The major limitations of learning-based algorithms are two-fold. First, they require designing specialized DNN models for different networking tasks, thus entailing high model engineering overhead~\cite{meng2020interpreting}. Second, they are prone to generalization issues, as DNNs may achieve poor performance on unseen data distributions/environments~\cite{kan2022improving}. \Revision{These problems, in fact, are not unique to networking and have been successfully addressed in other domains (e.g., NLP~\cite{kojima2022large,brown2020language} and robotics~\cite{rt22023zitkovich,driess2023palme}) by introducing LLMs as the foundation models to solve various downstream tasks. Hence, this work takes a pioneering step by systematically introducing LLMs into networking to address these problems.}

\begin{table*}[t]
	\renewcommand{\arraystretch}{0.6}
    \centering
    \caption{Information of three learning-based algorithm use cases in the networking area.}
    \label{tab:networking_tasks}
     \small
    \vspace{-0.3cm}
    \begin{tabular}{m{2.15cm}<{\raggedright}  m{5.1cm}<{\raggedright}  m{3.6cm}<{\raggedright}  m{3.9cm}<{\raggedright} m{1.4cm}<{\centering}}
    \toprule
    \textbf{Task} & \textbf{DNN Input} & \textbf{DNN Output} & \textbf{Objective} & \textbf{Learning Paradigm}\\
    \midrule
     Viewport Prediction (VP)    &  \textit{time-series:} historical viewports;\;\;\;\;\;\;\;\;\;\;\;\;\;\;\;\;\;\; \textit{image:} video content information & future viewports & minimize error between predicted and actual viewports & SL  \\
     \midrule
     Adaptive Bitrate Streaming (ABR) & \textit{time-series:} historical throughputs, delay; \textit{sequence:} chunk sizes at different bitrates; \;\;\;\;\;\;\;\;\;\;\;\;\;\;\;\;\;\;\;\;\;\;\;\;\;\;\;\;\;\;\;\;\;\;\;\;\;\;\;\;\;\;\;\;\;\;\;\;\textit{scalar:} current buffer length & bitrate selected for the next video chunk & maximize user's Quality of Experience (QoE) & RL \\
     \midrule
     Cluster Job Scheduling (CJS) & \textit{graph:} DAGs describing dependency and resource demands of job execution stages  & job stage  to run next, number of executors allocated to the stage & minimize job completion time & RL\\     
    \bottomrule
    \end{tabular}
\end{table*}

\subsection{Large Language Models}
\label{subsec:llm_bg}
The advent of large language models (LLMs) such as ChatGPT~\cite{chatgpt}, PaLM~\cite{aakanksha2023palm}, Llama2~\cite{touvron2023llama} and OPT~\cite{zhang2022opt} has profoundly revolutionized the field of natural language processing (NLP). These LLMs, pre-trained over large public corpora (e.g., wikis and books) to acquire extensive knowledge, have demonstrated remarkable capability across a wide range of applications that largely affect our daily life, including dialogue systems~\cite{glaese2022improving}, step-by-step reasoning~\cite{kojima2022large}, and even code generation~\cite{chen2021evaluating}. 

\begin{figure}[t]
\centering
\includegraphics[width=0.42\textwidth]{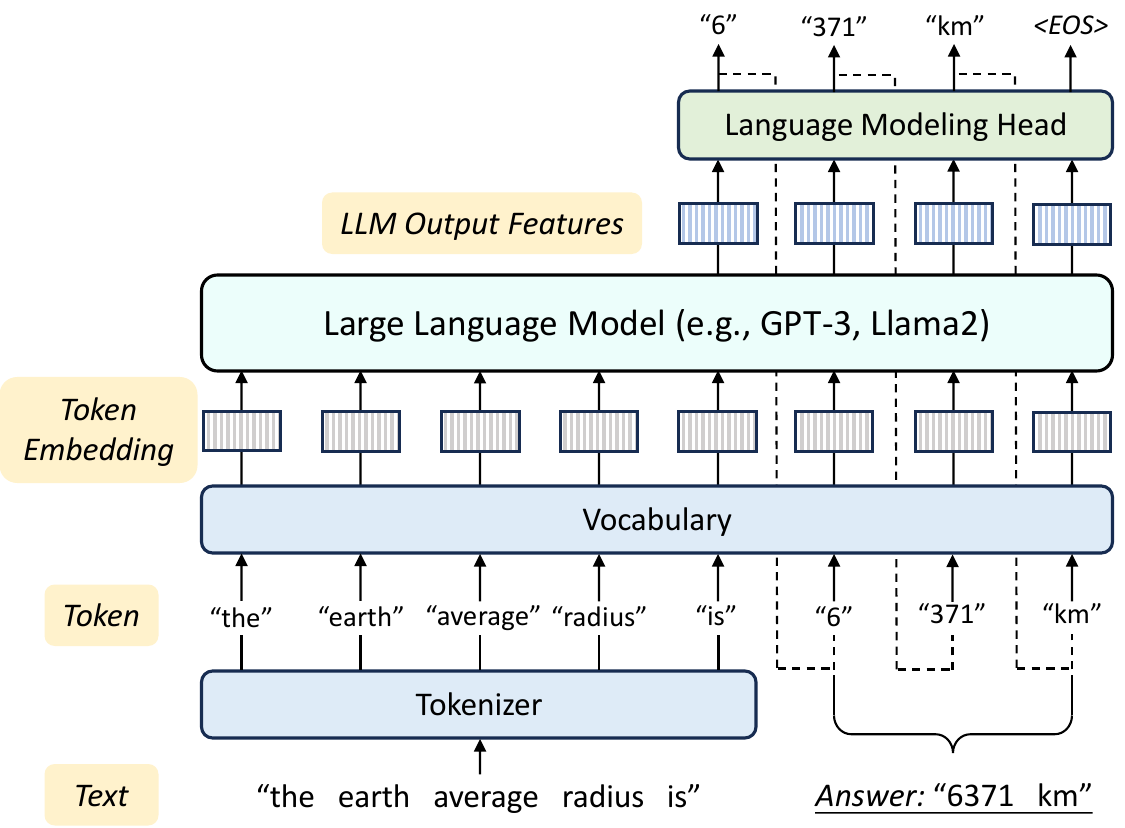}
\vspace{-0.2cm}
\caption{Illustration of the token-based answer generation mechanism of LLMs.}
\label{fig:llm_mechnism}
\vspace{-0.2cm}
\end{figure}

LLMs are essentially large DNNs built on top of Transformer~\cite{vaswani2017attention}, the de facto standard architecture for sequence modeling in NLP tasks. They model the inputs and outputs as sequences of tokens representing sub-words in NLP (e.g., a word “awesome” can be split into two tokens: “aw” and “esome”). Specifically, they take as input a sequence of tokens, and generate another sequence of tokens as answer with the assistance of three key components: \textit{tokenizer}, \textit{vocabulary} and \textit{language modeling (LM) head}. Figure~\ref{fig:llm_mechnism} illustrates the answer generation mechanism of LLM. Given an input sentence, the tokenizer splits it into a list of tokens. Then the vocabulary is used to map each token into an embedding vector that can be understood and effectively processed by LLM. Afterwards, LLM encodes these token embeddings into high-level features, which are subsequently passed to the LM head to predict the probability distribution of next token. Note that the output tokens are generated \textit{one by one} in the autoregressive manner~\cite{kwon2023efficient,liu2023cachegen}. Both the sequence of input tokens and previously generated tokens are repeatedly fed into the LLM to predict the next token, until an end-of-sentence token (\textit{<EOS>}) is emitted.

\subsection{Domain-Adapted LLMs}

The impressive performance of LLMs has sparked  pioneering research to adapt LLMs for other domains~\cite{driess2023palme,lin2023evolutionary,liu2023chipnemo,rt22023zitkovich}. For example, PaLM-E~\cite{driess2023palme} adapts PaLM~\cite{aakanksha2023palm} to generate step-by-step controlling commands for robotic manipulation. 
ESMFFold~\cite{lin2023evolutionary} showcases the successful applications of LLM in biological fields, which leverages LLM to predict atomic-level protein structure.

Inspired by these promising outcomes, this work explores the adaptation of LLMs for networking to address the limitations of existing learning-based algorithms, hoping to pave the way for more sustainable design philosophy of networking solutions. 
Unfortunately, although some prior studies have showcased that LLMs are capable of generating some technical documents for networking (e.g., description documents of digital twin for data centers~\cite{li2023chattwin}), none of the existing works provide an in-depth investigation on \textit{whether and how} LLMs can be adapted to solve networking tasks. Hence, this work tries to bridge this research gap and proposes an LLM adaptation framework for networking.

\section{Motivation}
\label{sec:motivation}

In this section, we identify the key challenges of LLM adaptation for networking, which motivate the design of \texttt{NetLLM}. We use the following three tasks (summarized in Table~\ref{tab:networking_tasks}) to make our discussion concrete:

\begin{itemize}[leftmargin=0.8em]

    \item  \textbf{Viewport prediction (VP)} serves as a fundamental building block of the emerging streaming systems of immersive videos (e.g., 360$^\circ$  videos~\cite{guan2019pano} and volumetric videos~\cite{han2020vivo}), where only the video content within the viewer's viewport (the region visible to viewer) is streamed in high quality to reduce the bandwidth consumption of video transmission~\cite{liu2023cav3,guan2019pano}. To accomplish this, the VP model predicts viewer's future viewport positions based on historical viewports, and potentially incorporates video content information (e.g., video frame) to enhance prediction performance~\cite{rondon2021track,wu2020spherical}. The goal of VP is to minimize the error between the predicted and viewer's actual viewports.

    \item \textbf{Adaptive bitrate streaming (ABR)} utilizes a RL model to dynamically adjust chunk-level bitrates  based on the perceived network conditions and playback buffer length during the streaming session of a video~\cite{xia2022genet,kan2022improving}. The objective of ABR is to maximize user's Quality of Experience (QoE), which is quantified by factors such as chunk bitrate, bitrate fluctuation, and rebuffering time.

    \item \textbf{Cluster job scheduling (CJS)} trains a RL scheduler to schedule incoming jobs within the distributed computing  cluster~\cite{mao2019learning,pen2021dl2}. 
    Each job is represented as a directed acyclic graph (DAG), which describes the dependency of each execution stage of the job and the resource demand of each stage.
    The primary task of RL scheduler is to select the next stage of a job to run and allocate a number of executors (computing resources) to that particular stage. The objective is to minimize the average job completion time, so that the system-level utilization of computing resources is optimized within the cluster. 
\end{itemize}
\noindent\textbf{Why these tasks?} We choose these tasks for several reasons.
\begin{itemize}[leftmargin=0.8em]
    \item First, they cover the two learning paradigms commonly adopted in networking, i.e., SL for prediction tasks (VP) and RL for decision-making tasks (ABR and CJS).
    \item Second, they include both centralized control (CJS) and distributed control (ABR) networking tasks. Specifically, the CJS scheduler is responsible for the entire cluster, while the ABR client independently selects bitrates without considering other clients.
    \item Finally, they involve diverse input modalities, covering the primary data modalities in many networking tasks. For example, many continuous signals in network adaptation problems (e.g., packet loss rate in congestion control~\cite{yen2023computers}) are represented as scalar data.
\end{itemize}
In particular, we choose VP as it encompasses multiple input modalities and requires cross-modality fusion, making it more challenging for LLM adaptation than other prediction tasks that generally involve single input modality (e.g., bandwidth prediction~\cite{mei2020realtime}). \textit{The characteristics of these tasks ensure that our subsequent discussion is representative and applicable to a wide range of networking scenarios.}

\begin{figure}[t]
    \centering
    \includegraphics[width=0.42\textwidth]{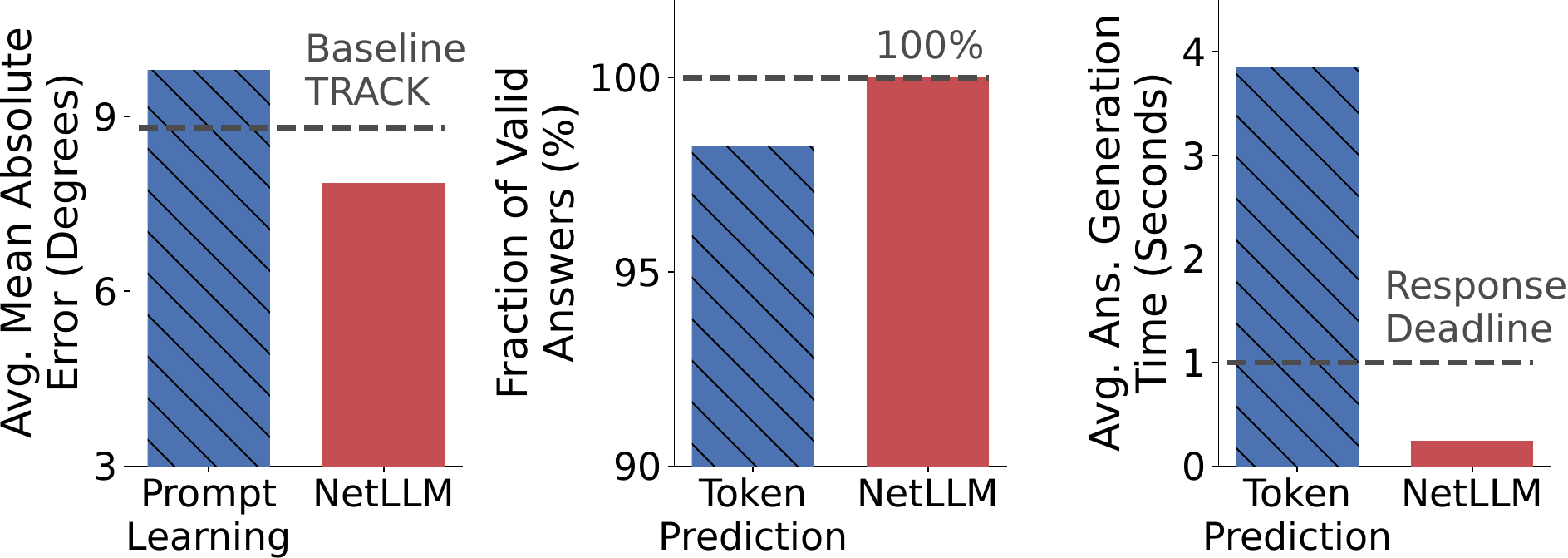}
    \vspace{-0.2cm}
    \caption{Illustration of the ineffectiveness for some natural alternatives with VP task as the example. \textit{Left}: Prompt learning~\cite{liu2023pre,min2023recent} that transforms data into textual prompts achieves sub-optimal performance, while \texttt{NetLLM} with a multimodal encoder to encode task input data effectively outperforms baseline. \textit{Middle, Right}: Token-based prediction with LM head fails to guarantee valid answers and produce stale responses, while  \texttt{NetLLM} efficiently addresses these issues with the networking head module.}
    \label{fig:challenge1and2}
    \vspace{-0.4cm}
\end{figure}

\noindent\textbf{Challenge 1: Large modality gap.} As shown in Table~\ref{tab:networking_tasks}, different networking tasks have different input information of diverse modalities, spanning from time-series network throughputs to DAG data. However, most LLMs are designed to accept plain text as inputs. Due to the substantial modality gap, it is impractical to directly feed the input information of networking tasks into LLMs for effective processing.

One seemingly natural approach to tackle this challenge is \textit{prompt learning}~\cite{liu2023pre,min2023recent,raffel2020exploring}, which transforms data into texts through a prompt template. Specifically, it designs a textual template that provides the information of task specifications, and uses this template to transform task inputs into textual prompts to instruct the LLM to generate desired answers that solve the tasks. While this approach shows promise in other fields, it falls short in the networking domain due to the following reasons. First, it is not feasible to transform data of complex modalities (e.g., image in VP and DAG in CJS) into textual prompts. Second, even if certain data can be converted into texts (e.g., time-series viewports in VP), we empirically observe that such transformation can be sub-optimal. To give a concrete example, we use this approach to adapt Llama2-7B~\cite{touvron2023llama} for the VP task. We design a template to encapsulate viewport data into prompts (we exclude video content information since it cannot be incorporated into prompts directly). Based on this prompt template, we instruct Llama2 to predict the future viewports in the next 1 second based on the historical viewports in the last 1 second (detailed setup of this measurement can be found in $\S$\ref{appendix:measurement_setup}). 

Figure~\ref{fig:challenge1and2} (\textit{left}) reports the performance of prompt-learning-adapted Llama2 in terms of mean absolute error (MAE). Lower MAE means better prediction performance. As shown, under the prompt learning framework, Llama2 achieves poor performance with 11.1\% higher MAE than the state-of-the-art VP model TRACK~\cite{rondon2021track}. This could be attributed to the significant modality gap between text and time-series data, as textual representation may not effectively express the characteristics of time-series data. For instance, the time-varying patterns  are typically not found in natural text.

\begin{figure}[t]
	\centering
	\includegraphics[width=0.42\textwidth]{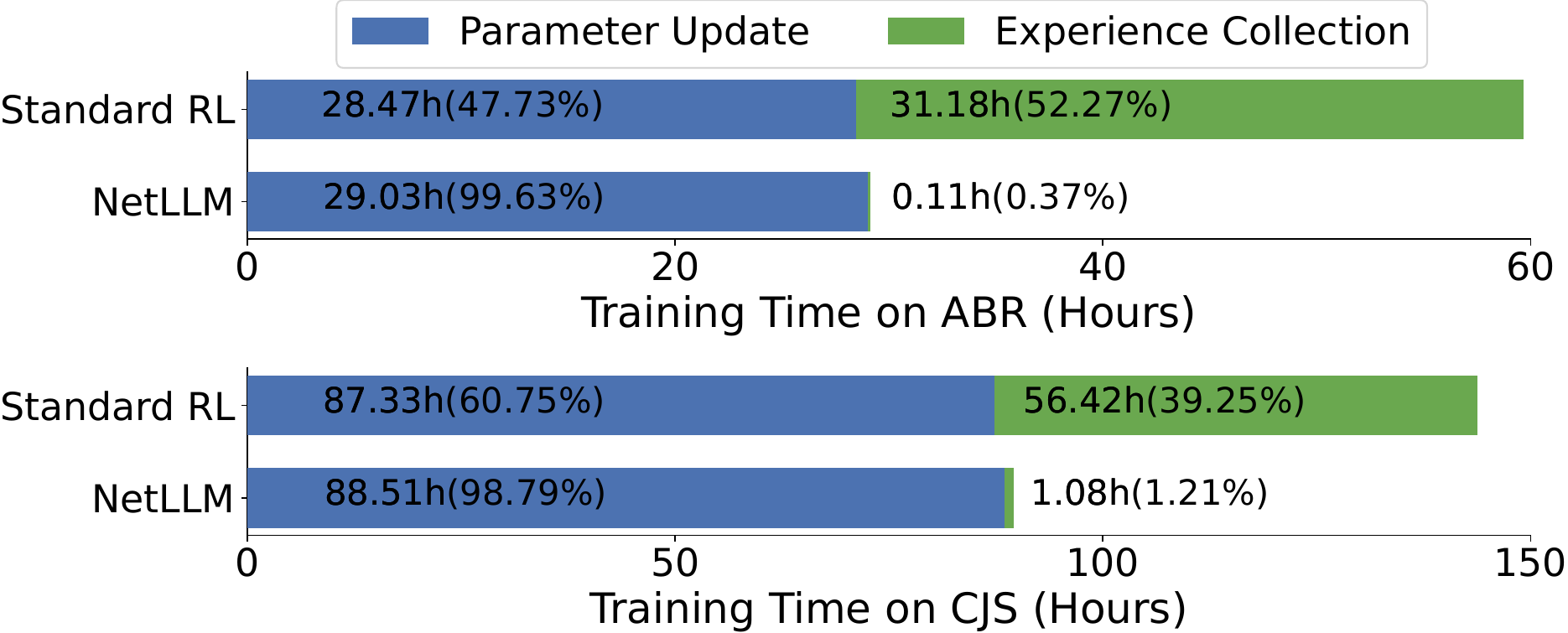}
	\vspace{-0.2cm}
	\caption{Using standard RL techniques~\cite{van2016deep,schulman2017proximal} to adapt LLM for RL-based decision-making tasks (ABR and CJS) incurs high training time due to the active environment interaction for experience collection. \texttt{NetLLM} eliminates this time-consuming process by designing an efficient data-driven adaptation pipeline in the DD-LRNA scheme.}
	\label{fig:std_rl_drawback}
	\vspace{-0.4cm}
\end{figure}

\begin{figure}[t]
	\centering
	\includegraphics[width=0.42\textwidth]{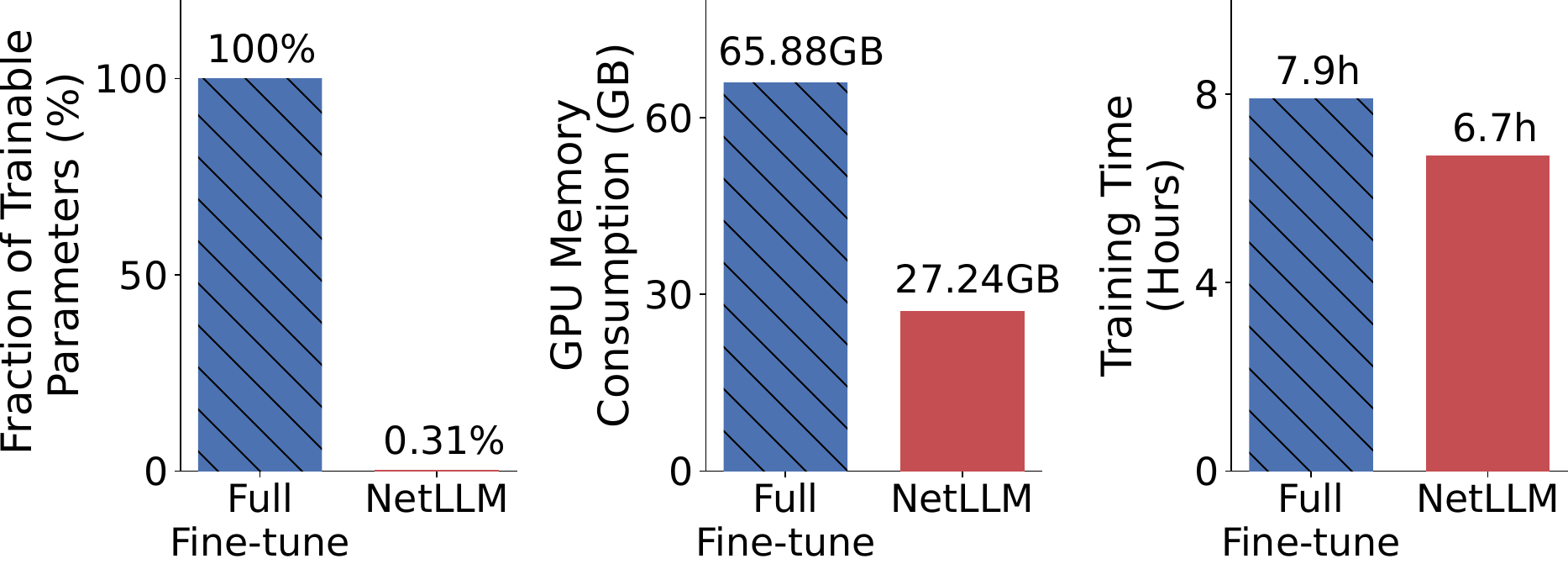}
	\vspace{-0.2cm}
	\caption{Illustration of the high adaptation costs of full-parameter fine-tune~\cite{touvron2023llama,chung2022scaling} on the VP task. The DD-LRNA scheme of \texttt{NetLLM} efficiently reduces the costs by introducing a set of small trainable low-rank matrices.}
	\label{fig:full_ft_drawback}
	\vspace{-0.4cm}
\end{figure}

\noindent\textbf{Challenge 2: Inefficiency of token-based answer generation.} As introduced in $\S$\ref{subsec:llm_bg}, by default, LLMs generate answers with LM head by predicting next tokens in an autoregressive manner. This, however, exhibits two main drawbacks in the networking area.

First, the uncertainty of token prediction increases the risk of LLM-generated answers to be physically invalid, a phenomenon known as \textit{hallucination}~\cite{ji2023survey,li2023halueval}.
To quantify this issue, we calculate the fraction of valid answers (see $\S$\ref{appendix:measurement_setup}) when adapting Llama2 for VP task based on token prediction. Figure~\ref{fig:challenge1and2} (\textit{middle}) shows that Llama2 fails to guarantee the generated answers to be 100\% valid when using token prediction. This raises concerns regarding the reliability of deploying LLMs for real-world network systems. 

Second, due to the sub-word nature of tokens, a single word may span multiple tokens. In consequence, LLMs often require multiple inferences to generate a single answer, as depicted in Figure~\ref{fig:llm_mechnism}. This  can produce delayed or even stale answers that fail to quickly adapt to the system changes. For instance, we measure the average time for Llama2 to generate a single answer for the VP task. Figure~\ref{fig:challenge1and2} (\textit{right}) shows that it takes up to 3.84s for per-answer generation, which significantly exceeds the 1-second response deadline required for predicting future viewports in the next second.

Note that the above problems are not unique to the networking area. Nevertheless, they reduce the efficiency of LLMs in networking, since networking tasks often require high reliability and quick responsiveness~\cite{meng2020interpreting,li2018auto}.

\noindent\textbf{Challenge 3: High adaptation costs.} Many networking tasks such as ABR~\cite{xia2022genet} and CJS~\cite{pen2021dl2} employ RL to solve complex system optimization problems, which involve active interaction between the environments (e.g., network environments with varying bandwidth or workload patterns) to collect experiences for reward optimization. In this context, simply fine-tuning LLMs for these tasks based on standard RL techniques (e.g., PPO~\cite{schulman2017proximal} and  DQN~\cite{van2016deep}) will introduce prohibitive adaptation costs due to the time-consuming process of environment interaction. 
To be more specific, we measure the amount of time of fine-tuning Llama2 over ABR (CJS) task for 10000 (100) iterations with standard RL. Each iteration involves interacting with the environment for one episode\footnote{An episode in RL refers to a single round of a RL model to interact with the environment from the initial state to final state. For example, in ABR task, the RL model starts streaming the first video chunk (initial state) and stops until all chunks are downloaded (final state).
} to collect experiences, followed by optimizing rewards based on these experiences. 
As depicted in Figure~\ref{fig:std_rl_drawback}, the experience collection caused by active environment interaction introduces \textit{additional} 31.18h (56.42h), accounting for 52.27\% (39.25\%) of the total training time on ABR (CJS) task.
While this problem has been observed in prior works~\cite{mao2017neural,mao2019learning}, it becomes intractable in the context of adapting LLMs for networking given their large parameter sizes. 

The adaptation costs become even more expensive when fine-tuning the full parameters of LLMs~\cite{ding2023parameter,fu2023on}. As shown in Figure~\ref{fig:full_ft_drawback}, fully fine-tuning Llama2-7B for the VP task consumes 65.88GB GPU memory and 7.9h of training time. This is because full-parameter fine-tune requires extensive memory and computation to store and maintain the training states due to the large parameter size. 
In fact, another practical drawback associated with full-parameter fine-tune is that it may disrupt the pre-trained knowledge of LLM since it updates all the parameters of LLM. As a result, this may prevent a single LLM to share across different networking tasks.

\begin{figure*}[t]
	\centering
	\includegraphics[width=0.85\textwidth]{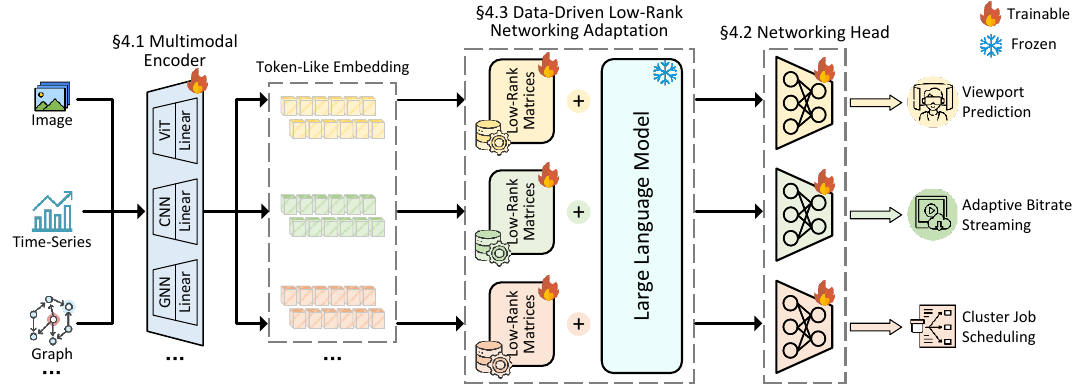}
	\vspace{-0.2cm}
	\caption{\texttt{NetLLM} consists of three core components: \textit{multimodal encoder} to encode task inputs, \textit{networking head} to generate task-specific answers and \textit{data-driven low-rank networking adaptation} to efficiently learn domain knowledge for networking. The framework is illustrated with three tasks: VP, ABR and CJS, but all ideas can be easily applied to other networking tasks.}
	\label{fig:framework}
	\vspace{-0.2cm}
\end{figure*}

\noindent\textbf{Summary.} In a nutshell, we observe three challenges of LLM adaptation for networking.
\begin{itemize}[leftmargin=0.8em]
   \item The large modality gap makes the input information of networking tasks incompatible with the LLM, preventing the LLM from effectively processing task inputs.
   \item The default token-based answer generation of the LLM exhibits inefficiency in the networking domain, which reduces the reliability and responsiveness for the LLM to serve network systems.
   \item The large parameter size of the LLM leads to significant costs to acquire domain-specific knowledge for networking, especially for RL-based tasks which require environment interaction. 
\end{itemize}

\section{NetLLM Design}
\label{sec:netllm_design}

In this section, we elaborate the detailed design of \texttt{NetLLM}, an innovative LLM adaptation framework for networking that efficiently solves the aforementioned challenges. As shown in Figure~\ref{fig:framework}, \texttt{NetLLM} comprises three main building blocks:
\begin{itemize}[leftmargin=0.8em]
    \item \textbf{Multimodal encoder.} \texttt{NetLLM} solves \textit{challenge 1} by designing an encoder module to encode multimodal input data of networking tasks into token-like embeddings, which can be effectively processed  by the LLM.
    \item \textbf{Networking head.} To address \textit{challenge 2}, \texttt{NetLLM} replaces the LM head for token prediction with different networking heads, which enable direct generation of a valid answer for specific  tasks in a single inference.
    \item \textbf{Data-driven low-rank networking adaptation (DD-LRNA).} To reduce the costs of adaptation, \texttt{NetLLM} develops an efficient DD-LRNA scheme to solve \textit{challenge 3}. It incorporates a data-driven pipeline to adapt LLMs for both prediction and decision-making tasks, and introduces different low-rank matrices to learn domain knowledge to further minimize the adaptation costs.
\end{itemize}
Note that during fine-tune, the parameters of the LLM are frozen to preserve pre-trained knowledge, while the multimodal encoder, networking heads, and low-rank matrices are tunable to optimize performance for different tasks. The details of each blocks are described as follows.

\begin{figure}[t]
	\centering
	\includegraphics[width=0.37\textwidth]{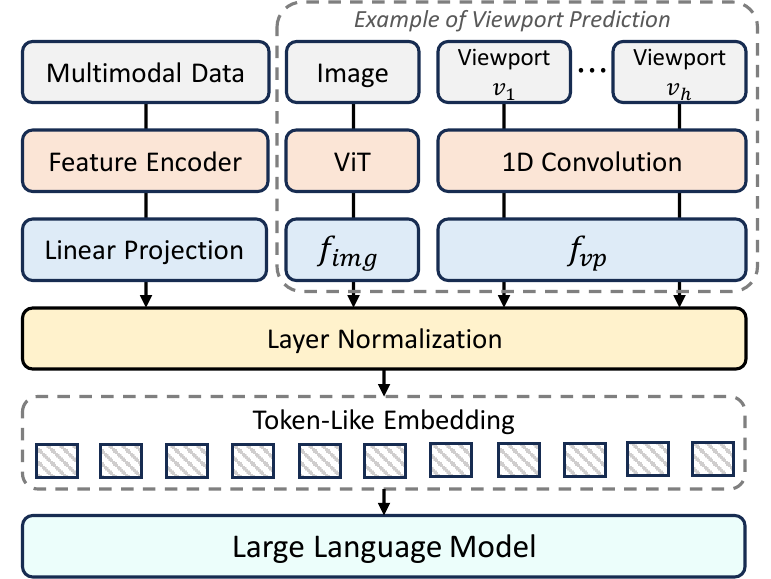}
	\vspace{-0.2cm}
	\caption{Illustration of the multimodal encoder of \texttt{NetLLM} to encode multimodal data.}
	\label{fig:multimodal_module}
	\vspace{-0.4cm}
\end{figure}

\subsection{Multimodal Encoder}

The key to processing task-specific information is to project the multimodal input data into token space to enable efficient utilization by the LLM. To achieve this, we design a multimodal encoder to automatically learn such projection. 
Figure~\ref{fig:multimodal_module} illustrates the  architecture of this module, which incorporates two blocks.

\noindent\textbf{Feature encoder.} We first employ different feature encoders to extract features from raw input data of various modalities. A key design consideration here is the choice of feature encoder for each modality. Instead of handcrafting feature encoders from scratch, which entails high model engineering costs, we reuse the existing well-designed encoders tailored for specific modalities. 
For example, Decima~\cite{mao2019learning} for CJS task develops a graph neural network (GNN)~\cite{wu2021comprehensive} encoder to extract features from DAG information, 
and Vision Transformer (ViT)~\cite{dosovitskiy2021image} has been widely used to encode images into hidden features. These designs are precious research outcomes and prove effective in processing specific modalities. Therefore, we cherish and efficiently utilize these designs by integrating them into our multimodal encoder module. 
\Revision{Specifically, we leverage the following feature encoders to encode different data modalities involved in networking: ViT for images, 1D-CNN for time-series and sequence data (e.g., historical throughputs and future chunk sizes at different bitrates in ABR), fully connected layer for scalar data (e.g., buffer occupancy in ABR), and GNN  for graph information (e.g., DAGs in CJS).}

\noindent\textbf{Linear projection.} The features extracted by encoders, however, may not align to the token space. For instance, features extracted by ViT have a dimension of 768~\cite{dosovitskiy2021image}, while Llama2 requires an input dimension of 4096~\cite{touvron2023llama}. To address this issue, we design a set of trainable linear layers to project the features extracted by different encoders. These layers automatically learn a highly efficient mapping from feature space to token space, producing a set of token-like embeddings that can be effectively utilized by the LLM. Additionally, we further enhance the projection process by normalizing the output embeddings with layer normalization~\cite{ba2016layer} to ensure training stability.

\noindent\textbf{Example.} Figure~\ref{fig:multimodal_module} illustrates the multimodal encoder with VP task as a concrete example. The ViT and 1D convolution layer (1D-CNN) are first used to encode the image and time-series viewport data, respectively. Next, the extracted features are projected into token-like embeddings with separate linear projection layers. Finally, all embeddings are normalized through layer normalization to ensure training stability, and passed to the LLM for further processing.
Figure~\ref{fig:challenge1and2} (\textit{left}) also provides statistical results to confirm the effectiveness of multimodal encoder to project task-specific input data. As shown, empowered by this module, \texttt{NetLLM} significantly outperforms prompt-learning based data processing scheme~\cite{liu2023pre}, with the average reduction of 19.7\% MAE for the VP task.

\subsection{Networking Head}
With the multimodal encoder, the LLM is capable to extract high-level features that encompass important task-specific information from input data of various modalities. These features are then fed into the networking head for direct answer generation. 
Specifically,  the networking head is designed as a trainable linear layer to predict task-specific answers based on LLM output features, which can be flexibly customized according to specific networking tasks. 
\Revision{Unlike LM head, the networking head constrains the answer generation into the valid range of possible answers, such as the range of valid viewport coordinates in VP and the set of video bitrates in ABR.}
In this way, all answers generated by LLM are guaranteed to be valid, ensuring the reliability of LLM for networking. Moreover, LLM can generate one answer within a single round of inference, thus significantly reducing the generation latency.

\begin{figure}[t]
	\centering
	\includegraphics[width=0.4\textwidth]{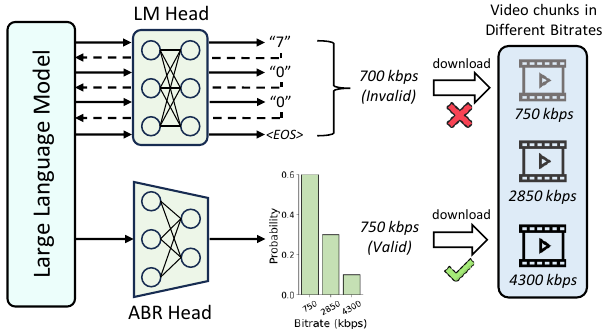}
	\vspace{-0.2cm}
	\caption{Comparison between LM head and networking head with ABR task as an example. For illustration, we assume that video chunks are encoded into three bitrate versions $\{750, 2850, 4300\}$ kbps. }
	\label{fig:task_head}
	\vspace{-0.4cm}
\end{figure}

\noindent\textbf{Example.} Figure~\ref{fig:task_head} compares the difference between LM head for token prediction and networking head with ABR task as the example. 
As depicted, the LM head generates answers by predicting next tokens autoregressively, which requires multiple rounds of inference and thus entails high generation latency.
Besides, due to the inherent uncertainty of token-based prediction, the generated answers may be invalid, such as a bitrate that does not exist. In contrast, the networking head is specially designed to predict the probability distribution of bitrates, enabling direct answer generation within a single round of inference \Revision{(e.g., the bitrate with the maximum probability is chosen as the answer in Figure~\ref{fig:task_head})}. Furthermore, since the outputs of networking head are limited to the discrete set of candidate bitrates, the generated answers are guaranteed to be always valid.
The superiority of networking head over LM head is also illustrated in Figure~\ref{fig:challenge1and2} (\textit{middle, right}), where \texttt{NetLLM} uses it to ensure the validness of answers and quickly produces answers before the response deadline for  VP task.

\subsection{Data-Driven Low-Rank Networking Adaptation}
In this part, we delve into the detailed design of the proposed data-driven low-rank networking adaptation (DD-LRNA) scheme to  efficiently fine-tune LLMs to acquire domain knowledge. The DD-LRNA comprises the following two core designs: i) a data-driven adaptation pipeline for both prediction and decision-making networking tasks; ii) a low-rank adaptation approach to constrain the fine-tuning process to a small number of parameters for more efficient adaptation.

\noindent\textbf{Data-driven networking adaptation.} In the case of prediction networking tasks, it is straightforward to fine-tune the LLM through the standard SL data-driven training pipeline. 
Specifically, given a task-specific dataset $\mathcal{D}_{sl} = \{\mathcal{X}, \mathcal{Y}\}$ of inputs $\boldsymbol{x}\in\mathcal{X}$ and labels $y\in \mathcal{Y}$, we leverage the multimodal encoder to encode input data $\boldsymbol{x}$, elicit prediction results $\hat{y}$ from the LLM with networking head, and compute loss for parameter update by:
\begin{equation}
\label{eq:sl_loss}
    \setlength{\abovedisplayskip}{3pt}
    L_{sl} = F_{sl}(y,\hat{y})
    \setlength{\belowdisplayskip}{3pt}
\end{equation}
where $F_{sl}$ is the loss function which can be cross entropy (CE) for classification tasks (e.g., traffic classification~\cite{pacheco2018towards}) or mean square error (MSE) for regression tasks (e.g., bandwidth prediction~\cite{mei2020realtime} and VP~\cite{rondon2021track}).

Nevertheless, when it comes to the decision-making tasks, the traditional RL training pipeline becomes impractical due to the time-consuming process of the interaction between the LLM and environments. 
To tackle this challenge, we design our RL adaptation pipeline based on the efficient data-driven RL techniques~\cite{prudencio2023survey,yen2023computers}, which tackle the same problem as traditional RL but without the need of environment interaction. \SecRevision{Specifically, we collect the experience dataset with \textit{any} existing (non-LLM) networking algorithms, and exploit this dataset to fine-tune the LLM for reward optimization. The intuition behind is to let the LLM learn a better policy by observing the behaviors of existing algorithms~\cite{levine2020offline}. That means, the LLM should not only learn from the good actions that lead to good performance, but also try to understand why some actions will lead to bad performance~\cite{yen2023computers}.} 

It is worth noting that, unlike traditional RL that requires periodic refreshing of the experience dataset during training~\cite{van2016deep,schulman2017proximal}, our approach allows the dataset to be collected \textit{only once }and used throughout the entire training process. As a result, the costs of adapting LLMs for RL-based networking tasks can be significantly reduced  (e.g., with 51.1\%/37.7\% reduction of training time for ABR/CJS task under the same training iterations, as depicted in Figure~\ref{fig:std_rl_drawback}).

\begin{figure}[t]
	\centering
	\includegraphics[width=0.42\textwidth]{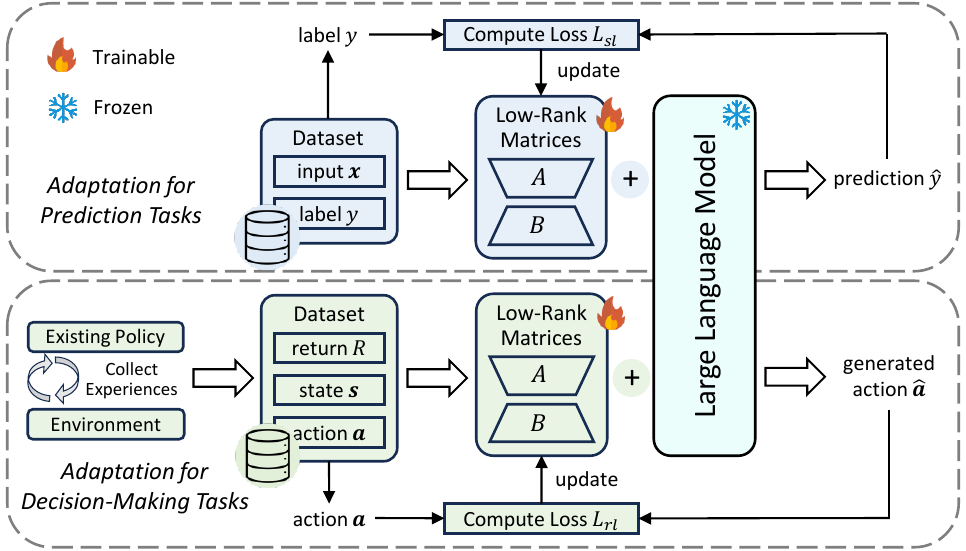}
	 \vspace{-0.2cm}
	\caption{Illustration of the data-driven low-rank networking adaptation scheme of \texttt{NetLLM}. }
	\label{fig:ddpeft}
	 \vspace{-0.4cm}
\end{figure}

The proposed RL adaptation pipeline is described as follows, which is built on top of the Transformer based data-driven RL~\cite{chen2021decision,janner2021offline} that caters to the sequence modeling nature of Transformer. Given a RL-based networking task, we first employ an existing policy (e.g, Decima~\cite{mao2019learning} for CJS) to collect an experience dataset which consists of experience trajectories: $\mathcal{D}_{rl}=\{\tau_1,\cdots,\tau_{|\mathcal{D}_{rl}|}\}$. Each trajectory $\tau=\{r_t,\boldsymbol{s}_t, \boldsymbol{a}_t\}_{t=1}^T$ is composed of rewards $r$, states $\boldsymbol{s}$ and actions $\boldsymbol{a}$, where $T$ denotes the episode length. For each sample in a trajectory $\{r_t,\boldsymbol{s}_t, \boldsymbol{a}_t\}\in \tau$, we substitute the reward $r_t$ by return $R_t=\sum_{i=t}^T r_i$ representing the cumulative rewards expected to receive from state $\boldsymbol{s_t}$. Additionally, considering that the state or action in some tasks may be constituted by multiple pieces of information (e.g., the state in ABR includes past network throughput and playback buffer length), we further discretize each state and action: $\boldsymbol{s}_t=\{s_t^1,\cdots,s_t^{n}\}, \boldsymbol{a}_t=\{a_t^1,\cdots, a_t^{m}\}$. This leads to the following representation of trajectory:
\begin{equation}
    \setlength{\abovedisplayskip}{3pt}
	\tau=\{R_t, s_t^1,\cdots,s_t^{n}, a_t^1,\cdots, a_t^{m}\}_{t=1}^T
    \setlength{\belowdisplayskip}{3pt}
\end{equation}

Based on the above trajectory representation, we then fine-tune the LLM to learn the distribution of returns. At each training step, we randomly sample a sequence of data from the dataset:
\begin{equation}
    \setlength{\abovedisplayskip}{3pt}
	d=\{R_i, s_i^1,\cdots,s_i^{n}, a_i^1,\cdots, a_i^{m}\}_{i=t-w+1}^t \in \mathcal{D}_{rl}
    \setlength{\belowdisplayskip}{3pt}
\end{equation}
where $w$ is the context window to facilitate the learning of return distribution. Next, we feed data $d$ to the LLM to generate actions $\{\hat{a}_i^1, \cdots, \hat{a}_i^m\}_{i=t-w+1}^t$. 
Notably, we consider return and each piece of state, action information as different modalities, and process them separately.
Finally, the training loss is calculated by:
\begin{equation}
\label{eq:rl_loss}
    \setlength{\abovedisplayskip}{3pt}
	L_{rl} = \frac{1}{w}\sum_{i = 1}^w \sum_{j=1}^m F_{rl}(a_i^j, \hat{a}_i^j)
    \setlength{\belowdisplayskip}{3pt}
\end{equation}
where $F_{rl}$ measures the difference between action $a_i^j$ and the generated action $\hat{a}_i^j$, which can be CE for discrete actions or MSE for continuous actions. 

The underlying rationale of the above training procedure is to train the LLM to model the distribution of returns conditioned on specific states, so that once trained, it can be used to generate a series of actions that achieve the desired returns~\cite{chen2021decision}. In particular, during the inference stage, we specify a target return based on the desired performance (e.g., maximum possible return to achieve excellent performance) to trigger the LLM to generate answers. 

\noindent\textbf{Low-rank networking adaptation.} 
With the data-driven adaptation pipeline in place, the LLM can now be fine-tuned for networking adaptation. 
Given the pre-trained parameters of the LLM denoted as $\Phi_0$, the goal of fine-tuning is to search for the parameter update $\Delta\Phi$ such that the resulting parameters $\Phi = \Phi_0 + \Delta\Phi$ are optimized for the specific networking task.
Nevertheless, due to the large parameter size of the LLM, directly fine-tuning the full parameters  entails prohibitive computation costs, as the dimension of learned parameters $|\Delta\Phi|$ is equal to $|\Phi_0|$ (e.g., $|\Phi_0| = 540$ billion for PaLM~\cite{aakanksha2023palm}).

To combat the above limitation, we freeze the parameters of LLM and introduce additional low-rank matrices to approximate the changes needed in the LLM parameters to learn domain-specific knowledge.
The underlying insight is that the parameter changes during adaptation (i.e., $\Delta\Phi$) reside on an intrinsic low rank~\cite{aghajanyan2020intrinsic,hu2021lora}.
Therefore, for each pre-trained matrix $W_0\in\Phi_0$ of dimension $d \times k$, we hypothesize the existence of a low rank $r\ll \min\{d, k\}$ and construct two low-rank matrices $A, B$ of dimension $d\times r, r\times k$ to approximate the update of $W_0$, i.e., $W= W_0 + \Delta W = W_0 + AB$. 
During adaptation, $W_0$ is frozen and all parameter update is constrained on matrices $A$ and $B$. 
As shown in Figure~\ref{fig:full_ft_drawback}, this significantly reduces the fine-tuning costs with the reduction of  60.9\% GPU memory and 15.1\% training time, since the low-ranks only introduces 0.31\% of trainable parameters.
Another benefit of this approach is that the pre-trained knowledge of the LLM is preserved as each $W_0\in\Phi_0$ is retained without any update. 
Hence, the same LLM can serve as the foundation model for different  tasks, and train different copies of $A, B$ to acquire different domain knowledge.

\noindent\textbf{Putting all together.} The DD-LRNA  scheme is briefly summarized in Figure~\ref{fig:ddpeft}. As shown, we freeze the parameters of LLM and allocate different trainable  low-rank matrices for each task. These matrices are then fine-tuned over a dataset to acquire domain-specific knowledge. For decision-making tasks, the dataset is collected by using existing algorithms to interact with environments. At each fine-tuning step, we sample a batch of data from the dataset, feed data to the LLM to generate answers, compute loss according to equation~(\ref{eq:sl_loss}) for prediction tasks or equation~(\ref{eq:rl_loss}) for decision-making tasks, and update the low-rank matrices through gradient descent. Note that in addition to low-rank matrices, the gradients are also propagated to update the parameters of multimodal encoder and networking head for performance optimization.

\subsection{Implementation}

\begin{figure}[t]
	\centering
	\includegraphics[width=0.42\textwidth]{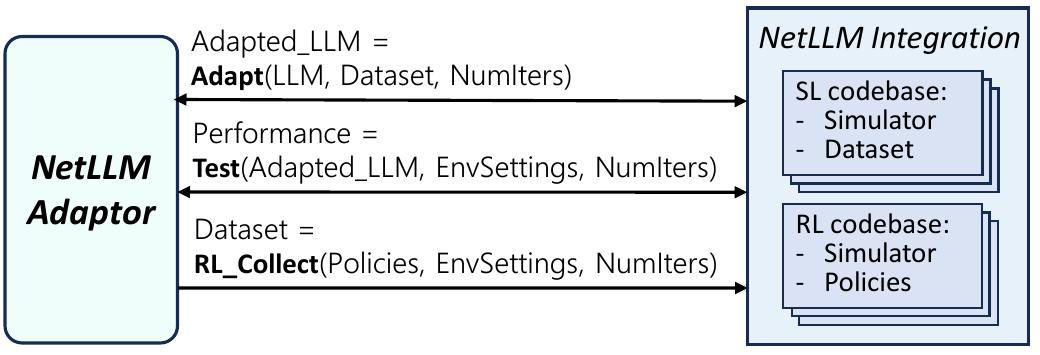}
	\vspace{-0.2cm}
	\caption{Components and interfaces needed to integrate \texttt{NetLLM} with an existing SL/RL codebase for LLM adaptation.}
	\label{fig:api}
	\vspace{-0.4cm}
\end{figure}

\texttt{NetLLM} is fully implemented in Python and Bash, and can be easily integrated into existing SL/RL codebases to adapt LLMs for networking tasks. As depicted in Figure~\ref{fig:api}, it interacts with an existing codebase with three APIs. First, \textit{Adapt} triggers \texttt{NetLLM} to use the provided dataset to adapt the LLM  to learn domain-specific knowledge for the target task, and returns the snapshot of the adapted LLM. Second, \textit{Test} evaluates the performance of the adapted LLM on the testing environments generated with the given simulation settings. Finally, for RL-based tasks without an available dataset for adaptation, \texttt{NetLLM} offers the \textit{RL\_Collect} API to collect the experience dataset by using the given RL policies to interact with the environments. Afterwards, the collected dataset can be plugged into the \textit{Adapt} API to adapt the LLM.

We have integrated \texttt{NetLLM} into three existing codebases for VP~\cite{wu2023mansy}, ABR~\cite{xia2022genet}, and CJS~\cite{decima_pytorch}, and implemented the above APIs based on the functionalities provided in the codebases. More details of implementation are provided in $\S$\ref{appendix:netllm_impl}.

\section{Evaluation}
\label{sec:evaluation}



\subsection{Setup}
\noindent \textbf{Simulation setup.} By default, we utilize Llama2-7B~\cite{touvron2023llama} as the foundation LLM. We then use \texttt{NetLLM} to adapt Llama2 for three networking tasks VP~\cite{wu2023mansy}, ABR~\cite{xia2022genet}, and CJS~\cite{mao2019learning}. We generate different simulation environments with real-world and synthetic datasets for training and testing, following the settings described in $\S$\ref{appendix:simulation_settings} and Table~\ref{table:vp_settings},~\ref{table:abr_settings},~\ref{table:cjs_settings}. These settings cover the key factors that affect the model performance. For instance, in ABR task, our environment settings consider the range and changing frequency of bandwidth as well as video bitrates.

\noindent \textbf{Baselines.} We implement three state-of-the-art learning-based algorithms for comparison: TRACK~\cite{rondon2021track} for VP, GENET \cite{xia2022genet} for ABR and Decima~\cite{mao2019learning} for CJS. We choose these baselines because they have open source implementation. In addition, we also compare with other rule-based (non-DNN) algorithms for each task: linear regression (labeled ``LR'')~\cite{qian2018flare} and velocity-based prediction (labeled ``Velocity'')~\cite{feng2021liveobj} for VP, BBA~\cite{huang2014buffer} and MPC~\cite{yin2015control} for ABR, first-in-first-out (labeled ``FIFO'') and fair scheduling (labeled ``Fair'')~\cite{spark_scheduler} for CJS. Appendix $\S$\ref{appendix:baseline_overview} provides the brief overview of all baselines. 

\noindent \textbf{Metrics.} For performance metrics, we consider mean absolute error (MAE) for VP, Quality of Experience (QoE) scores for ABR, job completion time (JCT) for CJS. Lower MAE, higher QoE and lower JCT indicate better performance. In particular, following the same formula in~\cite{mao2017neural,xia2022genet}, QoE is calculated as the weighted linear combination of bitrate, rebuffering time and bitrate changes. \SecRevision{Appendix $\S$\ref{appendix:eval_metrics} provides the details of three metrics.} 

\noindent \textbf{Hardware settings.} We conduct experiments  on a Linux server equipped with eight Intel(R) Xeon(R) Gold 5318Y CPUs and two NVIDIA 40GB A100 GPUs.

\begin{figure}[t]
		\centering
		\subfigure[Average performance with different random seeds]{
			\includegraphics[width=0.148\textwidth]{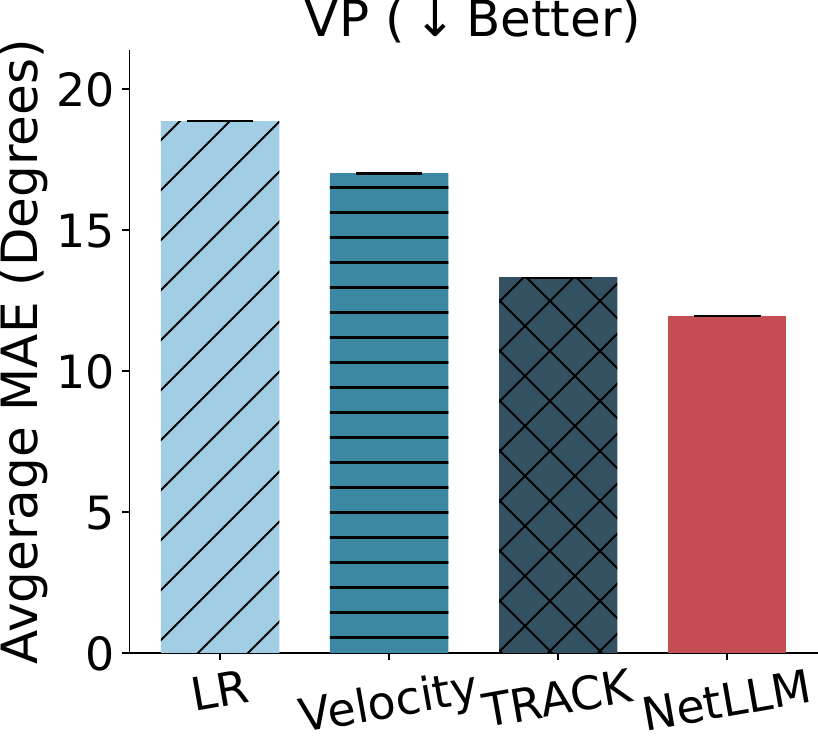}
			\hspace{0.05cm}
			\vspace{-0.2cm}
			\includegraphics[width=0.148\textwidth]{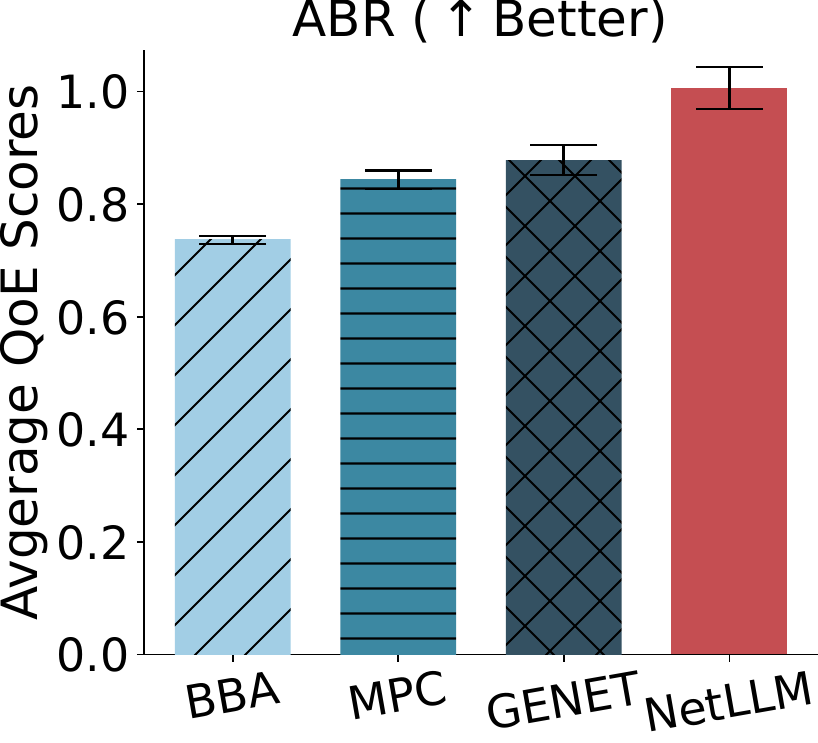}
			\hspace{0.05cm}
			\includegraphics[width=0.148\textwidth]{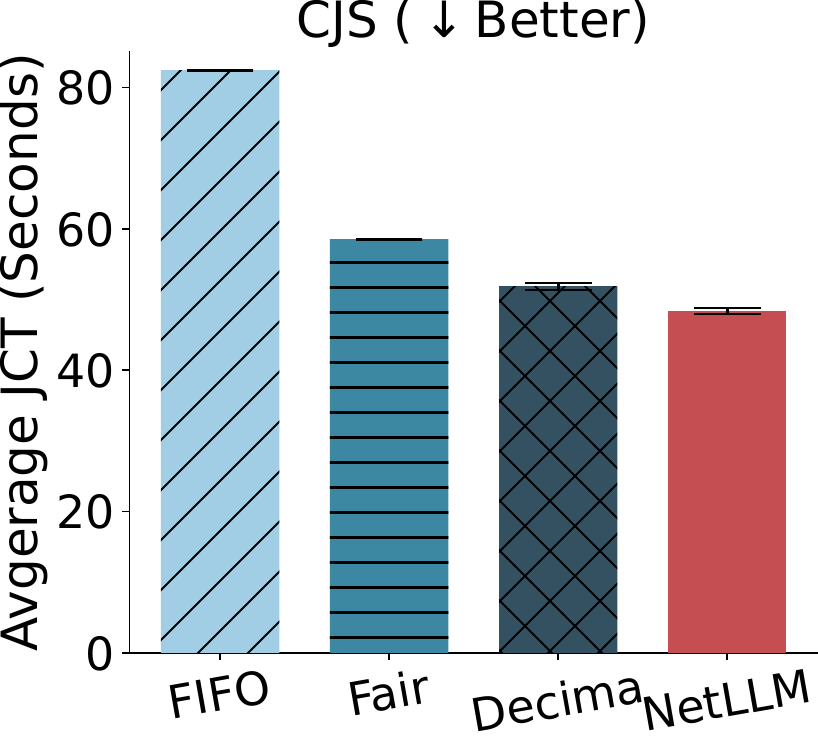}
		}
		\subfigure[CDF Performance]{
			\includegraphics[width=0.148\textwidth]{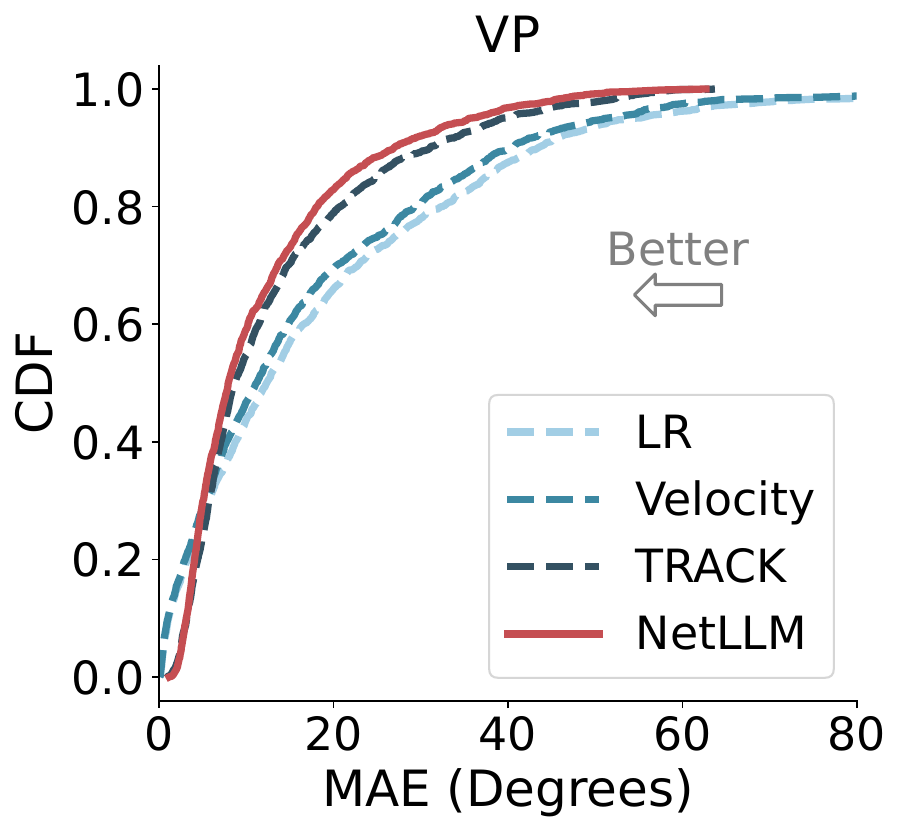}
			\hspace{0.05cm}
			\vspace{-0.4cm}
			\includegraphics[width=0.148\textwidth]{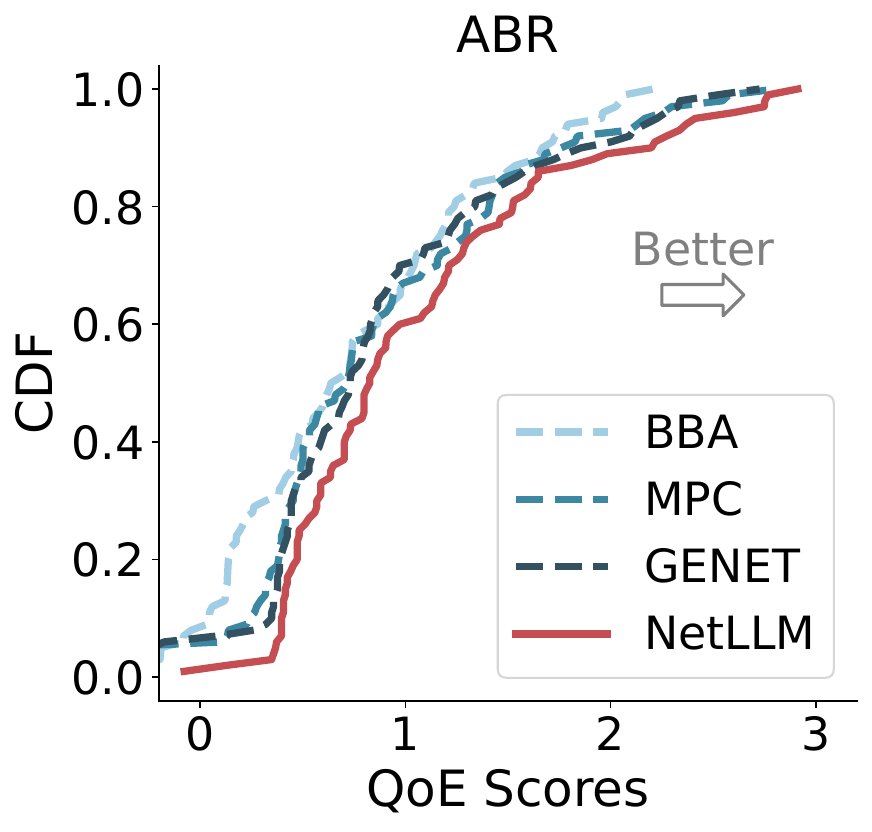}
			\hspace{0.05cm}
			\includegraphics[width=0.148\textwidth]{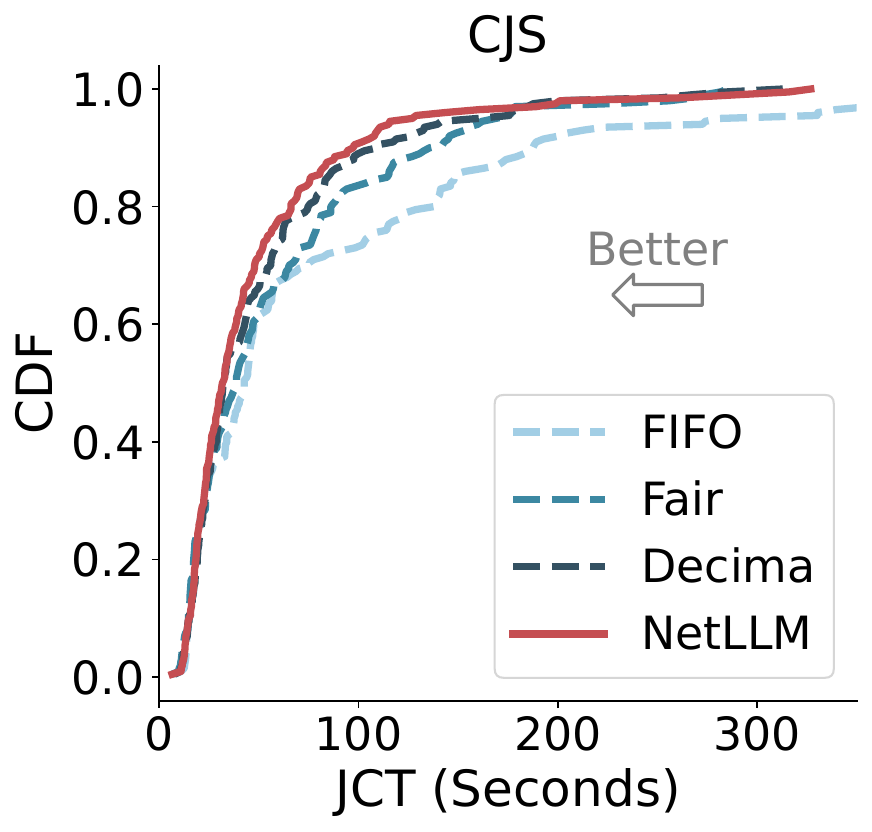}
		}
	\vspace{-0.3cm}
	\caption{Comparing \texttt{NetLLM}-adapted Llama2 for VP, ABR, and CJS, with baselines in testing environments generated with the same settings as training environments. }
	\label{fig:general_eval}
	\vspace{-0.4cm}
\end{figure}

\subsection{General Evaluation}
\label{subsec:general_eval}

In this part, we first compare \texttt{NetLLM}-adapted Llama2 with other methods across three different tasks over the testing environments with the same settings as training environments (see $\S$\ref{appendix:simulation_settings}). In other words, for each task, we adapt Llama2 and train learning-based algorithms over the environment generated with the target setting, and test all methods in the new environment from the same setting.

Figure~\ref{fig:general_eval} presents the performance of each method for the corresponding tasks. 
As shown in Figure~\ref{fig:general_eval}, \texttt{NetLLM}-adapted Llama2 consistently outperforms other methods across all cases.
It surpasses all baselines  by reducing 10.1-36.6\% of MAE for VP, improving 14.5-36.6\% of QoE for ABR and reducing 6.8-41.3\% of JCT for CJS. 
Figure~\ref{fig:general_eval} also provides more detailed results in the form of CDF for each task. It can be seen that a large proportion of \texttt{NetLLM}-adapted Llama2 is concentrated in the range of lower MAE, higher QoE and lower JCT. For instance, for CJS task, the 90th percentile JCT of Llama2 is 97.3 seconds, while this value dramatically increases to 109.3 seconds for Decima, 135.6 seconds for Fair and 187.5 seconds for FIFO. The above outcomes highlight the effectiveness of \texttt{NetLLM} in LLM adaption for networking.

\begin{figure}[t]
	\centering
	\subfigure[VP ($\downarrow$ Better)]{
		\centering
		\includegraphics[width=0.148\textwidth]{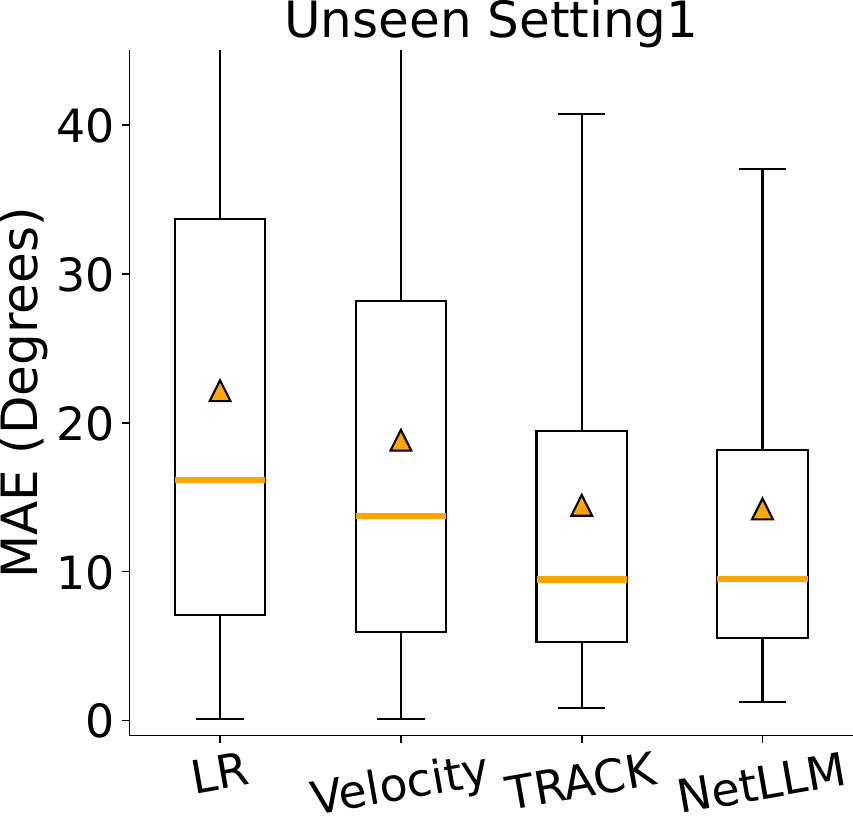}
		\hspace{0.05cm}
		\includegraphics[width=0.148\textwidth]{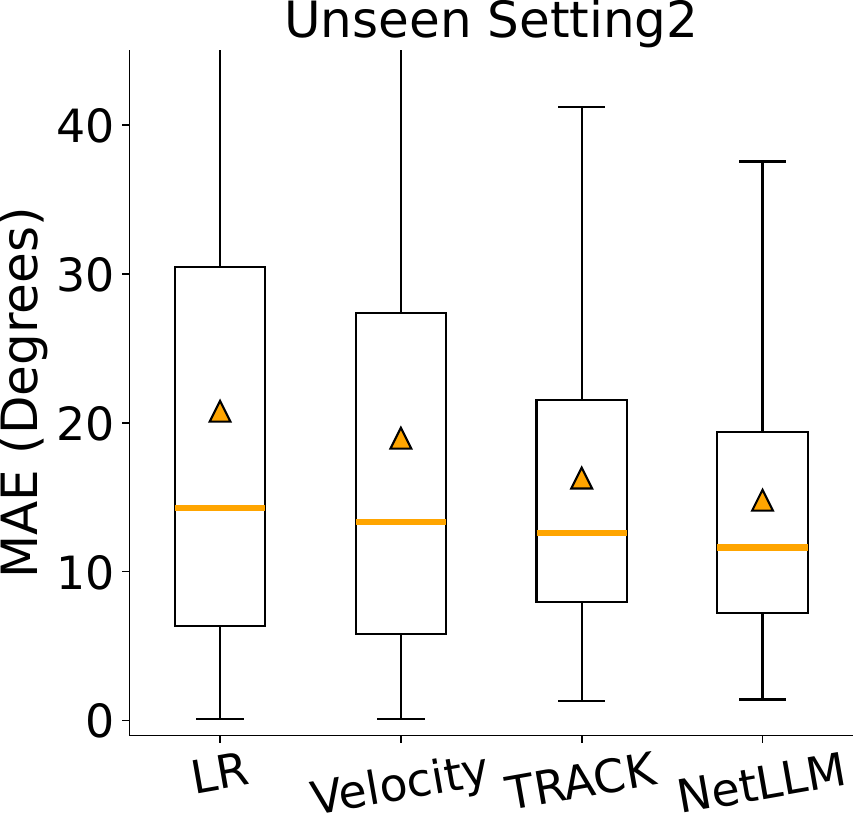}
		\hspace{0.05cm}
		\includegraphics[width=0.148\textwidth]{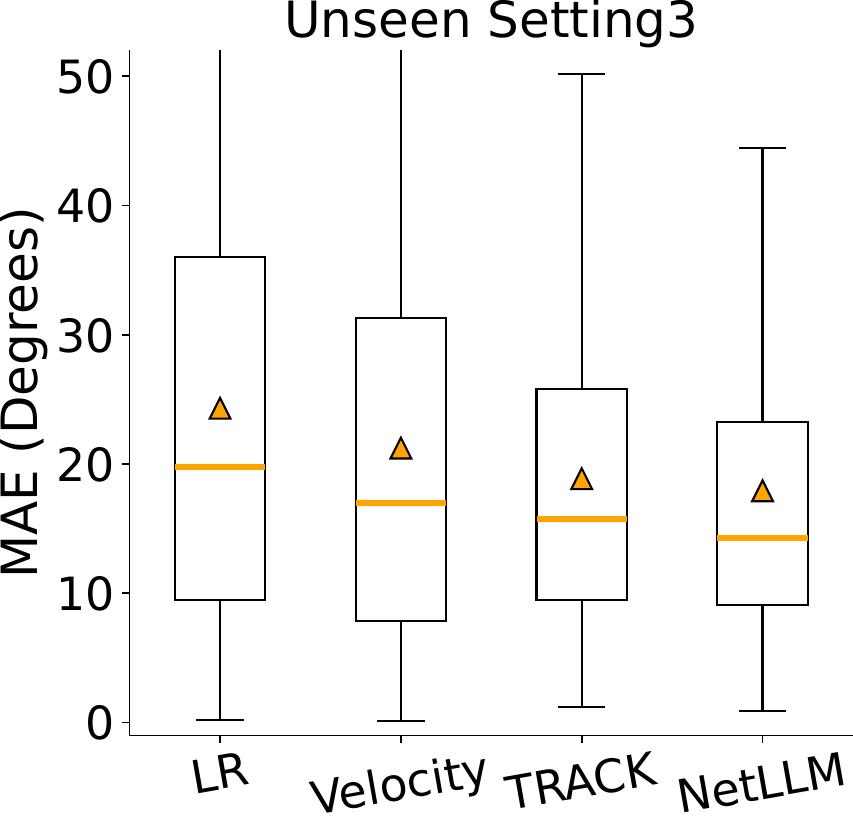}
	}
	\subfigure[ABR ($\uparrow$ Better)]{
		\centering
		\includegraphics[width=0.148\textwidth]{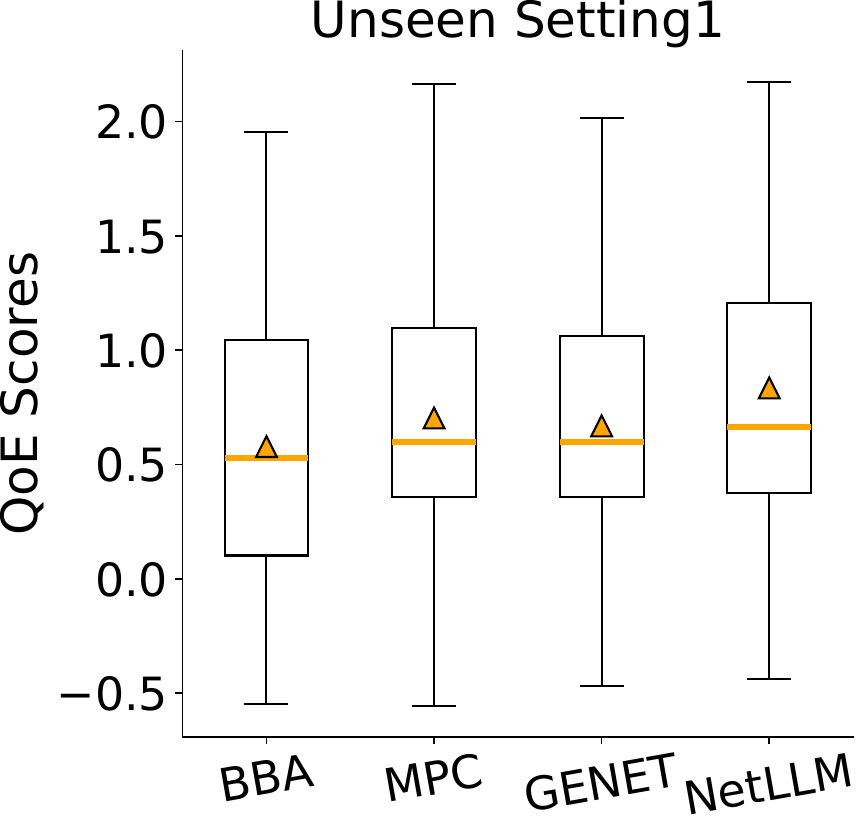}
		\hspace{0.05cm}
		\includegraphics[width=0.148\textwidth]{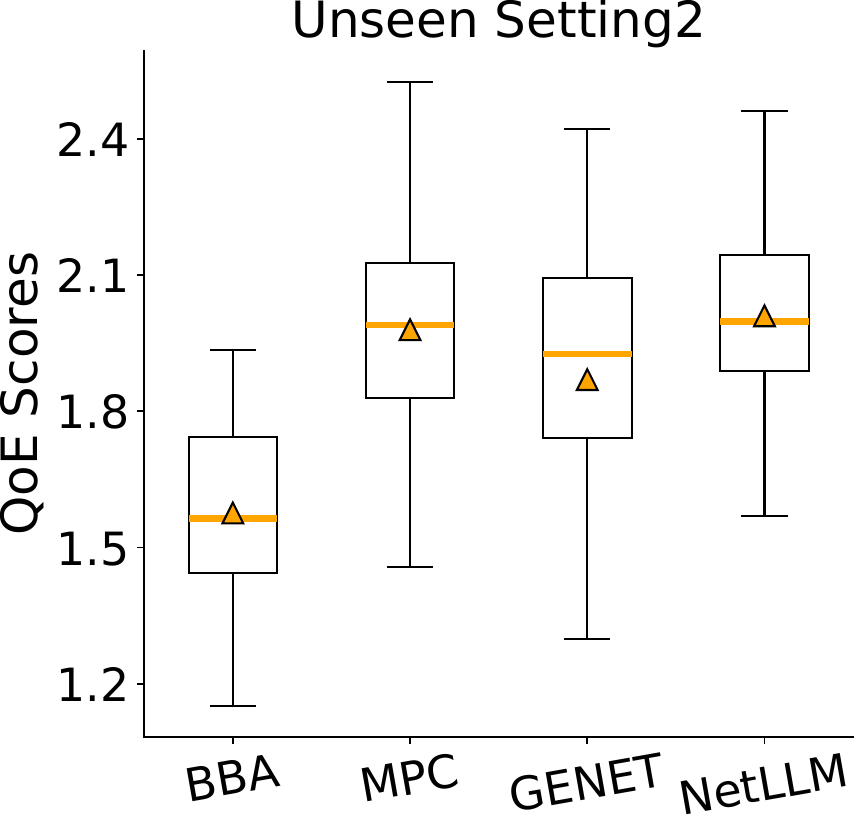}
		\hspace{0.05cm}
		\includegraphics[width=0.148\textwidth]{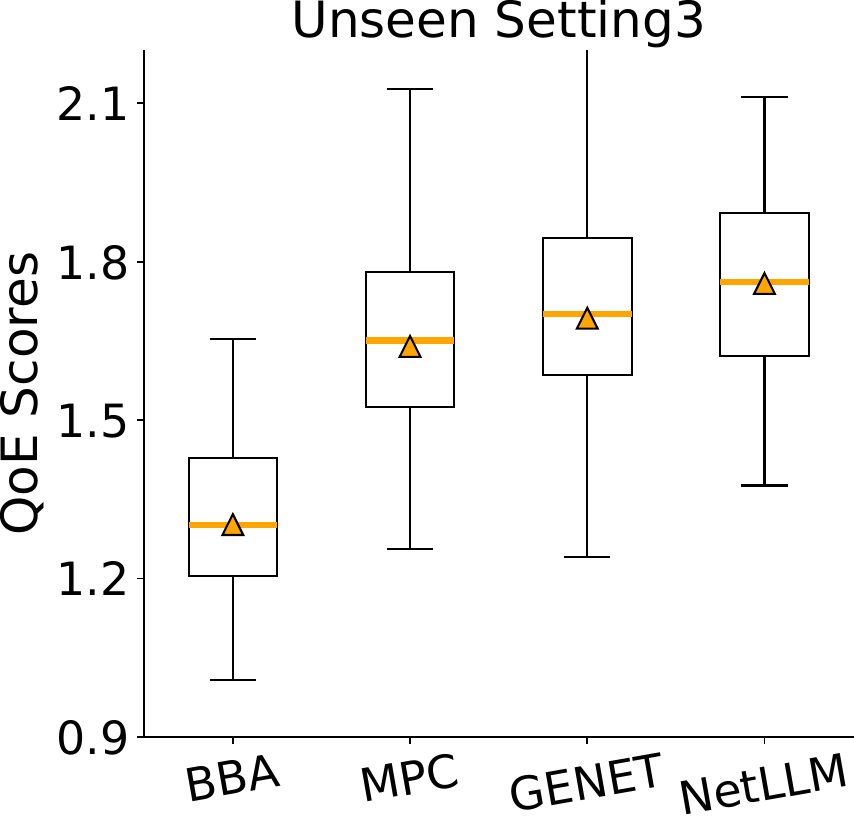}
	}
	\subfigure[CJS ($\downarrow$ Better)]{
		\centering
		\includegraphics[width=0.148\textwidth]{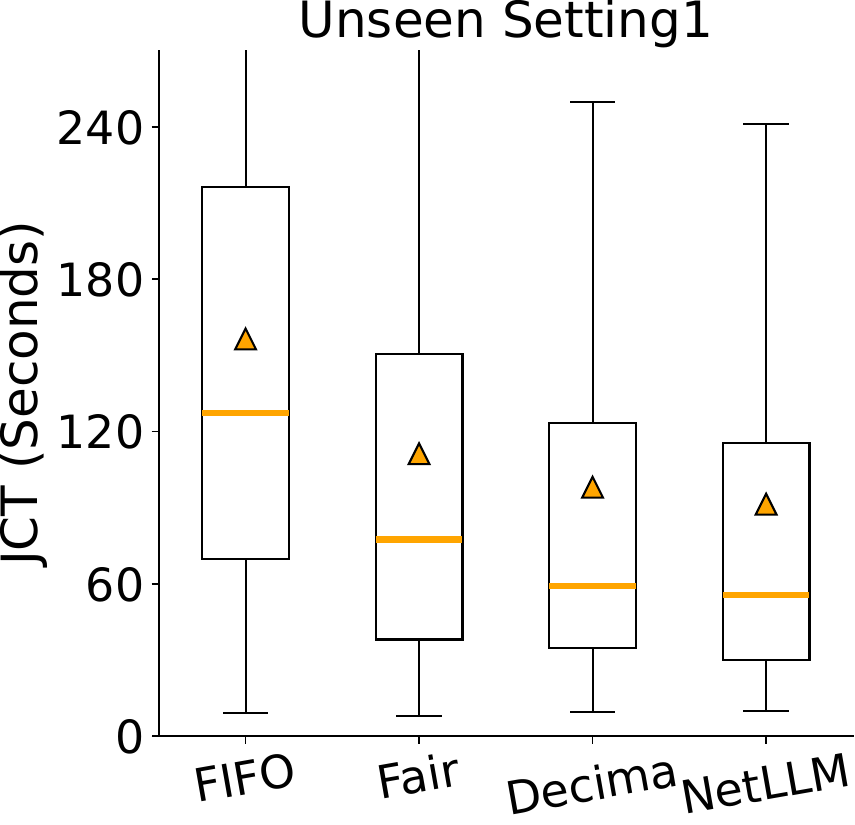}
		\hspace{0.05cm}
		\includegraphics[width=0.148\textwidth]{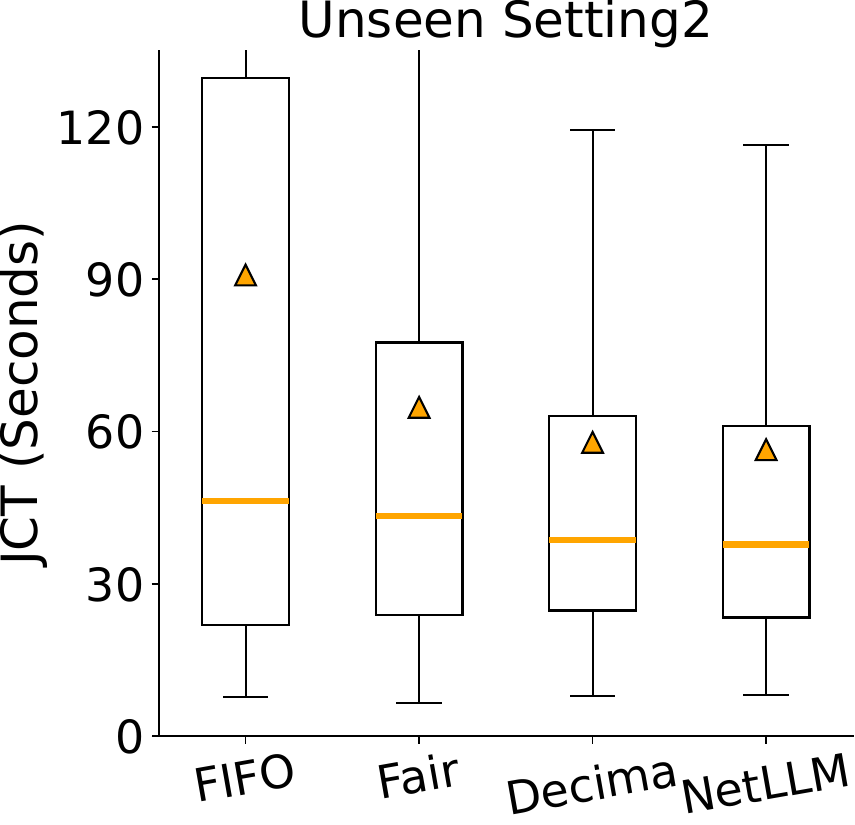}
		\hspace{0.05cm}
		\includegraphics[width=0.148\textwidth]{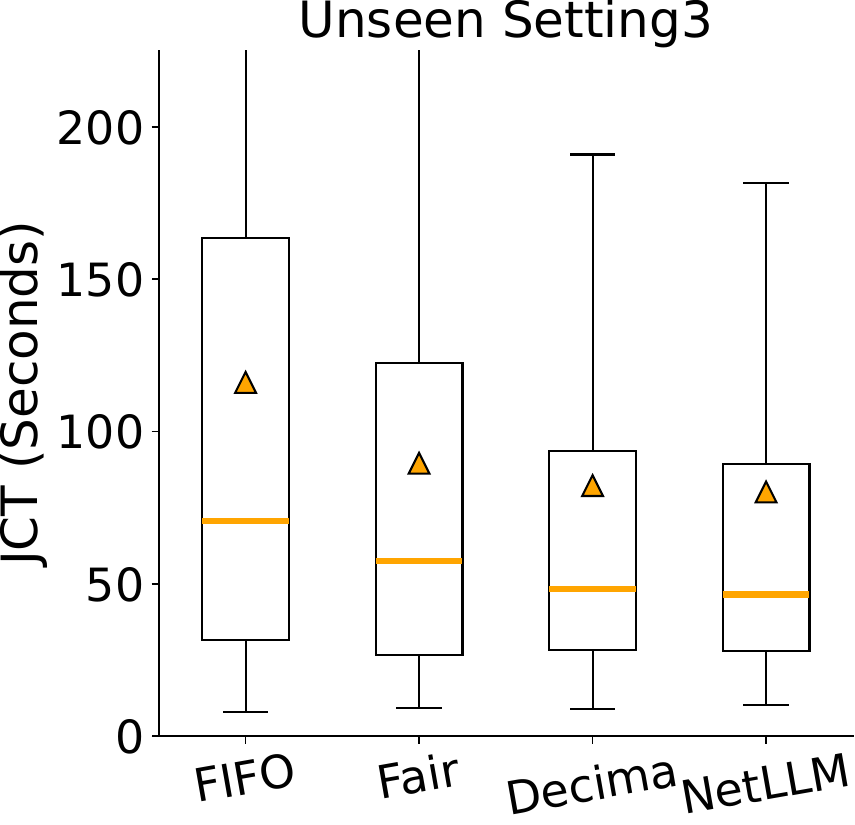}
	}
	\vspace{-0.3cm}
	\caption{Comparing the generalization performance of \texttt{NetLLM}-adapted Llama2 for VP, ABR, and CJS, with baselines in testing environments generated with settings different from training environments. The shape of box shows the distribution and the triangle in each box denotes average.}
	\label{fig:generalization}
	\vspace{-0.4cm}
\end{figure}

\begin{figure*}[t]
	\centering
	\subfigure[ABR Unseen Setting 1]{
		\includegraphics[width=0.3\textwidth]{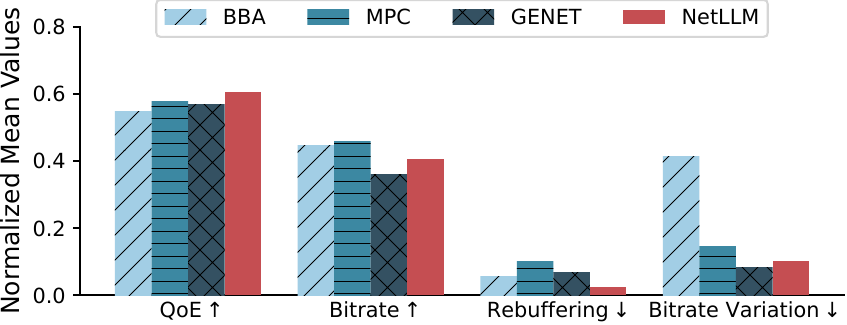}  
	}
	\hspace{0.05cm}
	\subfigure[ABR Unseen Setting 2]{
		\includegraphics[width=0.3\textwidth]{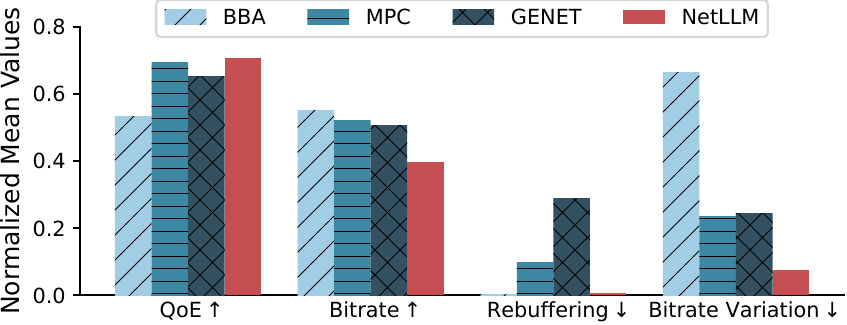}
	}
	\hspace{0.05cm}
	\subfigure[ABR Unseen Setting 3]{
		\includegraphics[width=0.3\textwidth]{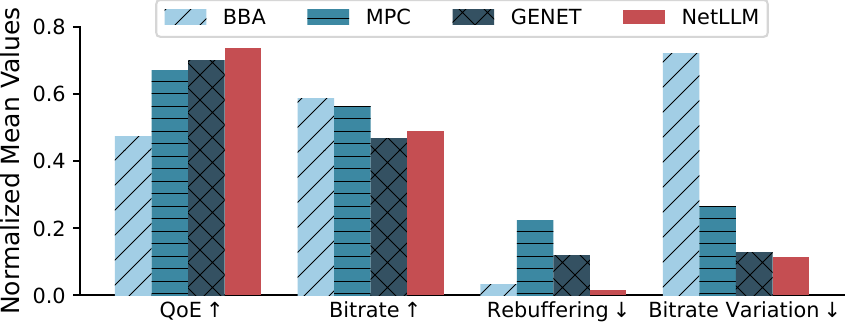}
	}
	\vspace{-0.3cm}
	\caption{Comparing \texttt{NetLLM}-adapted Llama2 with baselines for ABR by breaking down their performance on individual QoE factors in different unseen environments. Results are normalized through min-max. Arrow $\uparrow$ / $\downarrow$ means higher/lower is better.}
	\label{fig:qoe_breakdown}
	\vspace{-0.4cm}
\end{figure*}

\SecRevision{It can be seen that there is a clear ranking among the three baselines on each task, where the learning-based algorithms consistently yield improvement over traditional rule-based algorithms. This can be attributed to the inherent strength  of DNNs in fitting complex functions for prediction. However, the LLM demonstrates even more powerful  capabilities in function approximation, pattern mining and long-term planning, owing to its large parameter size and large-scale pre-training. Hence, by effectively utilizing the strengths of LLM to solve networking tasks, \texttt{NetLLM} achieves superior performance compared to the other learning-based algorithms. In addition, it is worth mentioning that the performance gain of learning-based algorithms relies on engineering specialized DNN models for the target tasks. In contrast, \texttt{NetLLM} efficiently utilizes the LLM as the foundation model for task solving. That said, it uses the same LLM to solve various networking tasks without any further modification on the model, thus significantly reducing the overhead of model engineering.}

\subsection{Generalization}
\label{subsec:generalization}

Next, we evaluate the generalization performance of all methods for each task in testing environments generated with various settings different from the training environments (see $\S$\ref{appendix:simulation_settings}). As depicted in Figure~\ref{fig:generalization}, \texttt{NetLLM}-adapted Llama2 consistently outperforms baselines in terms of average values and distributions across all cases. For instance, compared to the learning-based algorithms, it reduces the MAE by 1.7-9.1\%, improves the QoE by 3.9-24.8\% and reduces the JCT by 2.5-6.8\% on average.
This suggests that, enabled by \texttt{NetLLM}, Llama2 demonstrates superior generalization performance. 

From Figure~\ref{fig:generalization}, we also notice that learning-based algorithms do not always outperform conventional rule-based algorithms for the ABR task. 
Figure~\ref{fig:qoe_breakdown} breaks down the QoE scores of all ABR methods for more detailed analysis. 
As shown, GENET is surpassed by MPC with 5.2\%/5.9\% lower average QoE on unseen setting 1/2. More specifically, on unseen setting 1, where the streaming video is different from training one, GENET fails to optimize video bitrates and thus achieves worse performance than MPC. 
\SecRevision{On the other hand, GENET struggles to adapt to the dynamic fluctuations of bandwidth on unseen setting 2, where the testing bandwidth traces change more frequently than training ones. It may inappropriately select high bitrates when the current bandwidth resources become scarce, and thus produces the highest rebuffering time among other methods.} 
In contrast, \texttt{NetLLM}-adapted Llama2 strikes a good balance between the three QoE factors and thus achieves the highest QoE scores on all settings. These cases exemplify that conventional DNN models may perform poorly in unseen environments. In comparison, leveraging our \texttt{NetLLM} framework, we can indeed efficiently utilize the extensive knowledge of the LLM to achieve stronger generalization.

\begin{figure}[t]
	\centering
	\includegraphics[width=0.35\textwidth]{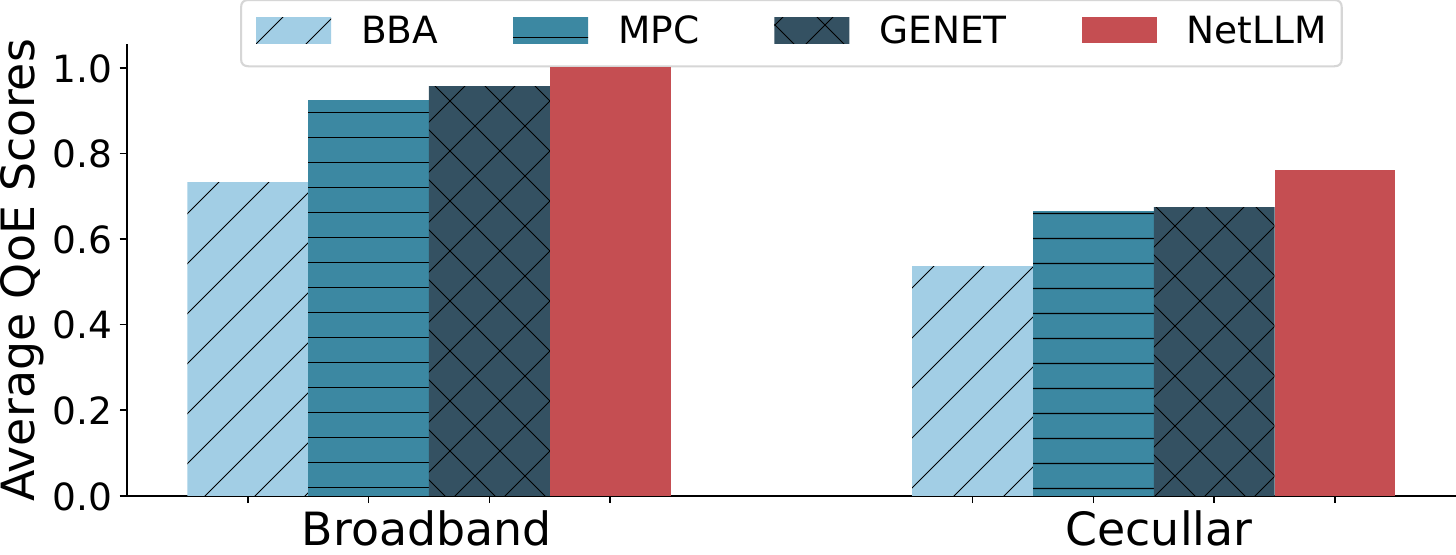}
	\vspace{-0.2cm}
	\caption{Comparing \texttt{NetLLM}-adapted Llama2 with baselines for ABR on real-world environments with different network connections.}
	\label{fig:real_world}
	\vspace{-0.4cm}
\end{figure}

\textbf{Real-world tests.} As a final test of generalization, we evaluate  \texttt{NetLLM}-adapted Llama2 in a real-world client-server ABR system under different network connections (see $\S$\ref{appendix:real_world} for detailed setup). The results are reported in Figure~\ref{fig:real_world}. On each network, the adapted Llama2 outperforms the baselines. This indicates that the LLM adapted by \texttt{NetLLM} is able to generalize to real-world scenarios.

\begin{figure}[t]
	\centering
	\includegraphics[width=0.44\textwidth]{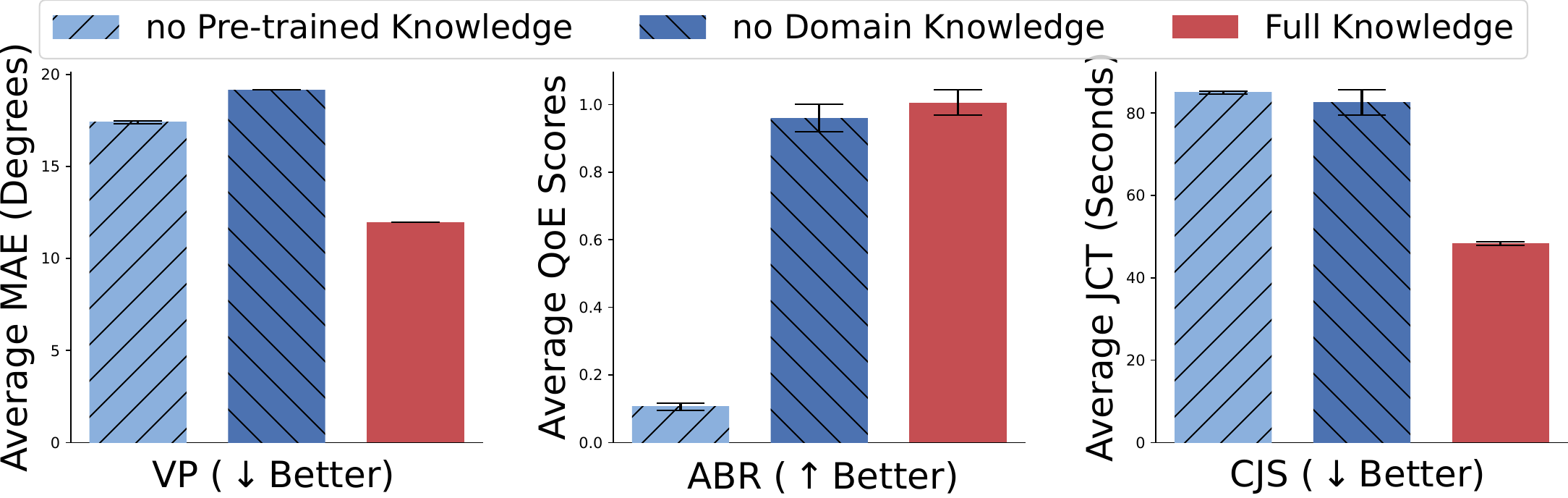}
	\vspace{-0.2cm}
	\caption{Exploring the importance of pre-trained and learned domain-specific knowledge of LLM in networking adaptation.}
	\label{fig:importance_knowledge}
	\vspace{-0.4cm}
\end{figure}

\subsection{Deep Dive}
\label{subsec:deep_dive}

\noindent\textbf{Importance of pre-trained and domain knowledge.} To gain a deeper understanding of why LLMs can be adapted for networking, we investigate the importance of both the pre-trained and learned domain knowledge of LLMs in networking adaptation. We use Llama2-7B as the LLM for our exploration. First, we disable the pre-trained weights of Llama2 that represent its pre-trained knowledge, randomly initialize its weights and train it from scratch for each task. As depicted in Figure~\ref{fig:importance_knowledge}, the absence of pre-trained knowledge leads to dramatic performance drop across all tasks. 
\Revision{This indicates that while LLMs are pre-trained over text corpora to acquire language knowledge, their emergent abilities (e.g., planning~\cite{xi2023rise}, pattern mining~\cite{rt22023zitkovich}) are indeed universal and applicable across domains, including networking. For instance, the pattern mining ability of LLM can be utilized to mine complex changing patterns of viewport movement for accurate viewport prediction. Hence, the pre-trained knowledge of LLMs is crucial for networking adaptation.}

Next, we preserve the pre-trained knowledge of Llama2 but disable the low-rank matrices that represent the learned domain knowledge. 
As reported in Figure~\ref{fig:importance_knowledge}, the absence of domain knowledge also results in performance degradation on each task, which highlights the importance of \texttt{NetLLM} to acquire domain knowledge.

\noindent\textbf{Impacts of different types of LLMs.}
To validate whether \texttt{NetLLM} is applicable to various LLMs, 
we employ it to adapt three additional LLMs besides Llama2 for the VP and ABR tasks: OPT~\cite{zhang2022opt}, Mistral~\cite{jiang2023mistral} and LLaVa~\cite{liu2023visual}. The size of each LLM is set to 7B for fair comparison. It is worth mentioning that LLaVa is a multimodal LLM trained on the combination of image and text corpora. We select LLaVa for our evaluation to investigate whether the pre-trained knowledge of multimodal fusion is applicable for networking.

\begin{figure}[t]
    \centering
    \includegraphics[width=0.44\textwidth]{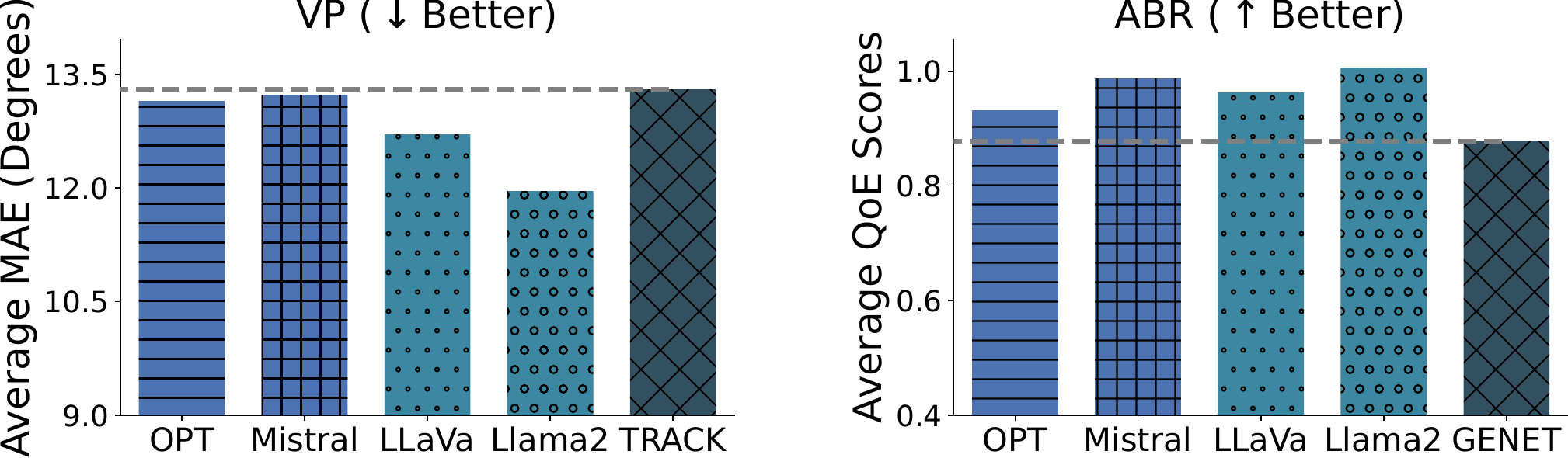}
    \vspace{-0.2cm}
    \caption{Comparing the performance of different LLMs adapted by \texttt{NetLLM} for VP and ABR with  learning-based algorithms.}
    \label{fig:different_llm}
    \vspace{-0.4cm}
\end{figure}

As shown in Figure~\ref{fig:different_llm}, all the adapted LLMs outperform the state of the arts on both tasks, which confirms the compatibility of \texttt{NetLLM}.
Interestingly, we observe that  the multimodal LLaVa performs worse than to single-modal Llama2. This suggests that the knowledge acquired by LLaVa in multimodal fusion during pre-training may not be directly beneficial in the networking domain\footnote{We leave deeper investigation of the efficacy of multimodal LLMs in networking for future work.}.

\begin{figure}[t]
    \centering
    \includegraphics[width=0.44\textwidth]{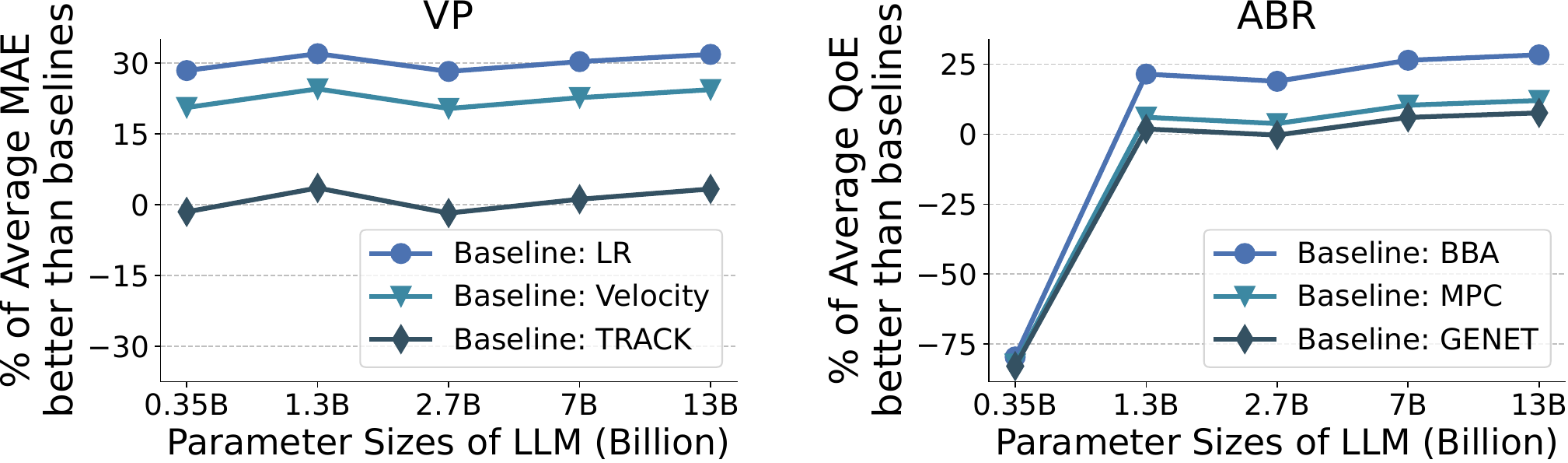}
    \vspace{-0.2cm}
    \caption{Exploring the impacts of LLM sizes in networking adaptation, with OPT~\cite{zhang2022opt} as the foundation model.}
    \label{fig:llm_sizes}
    \vspace{-0.4cm}
\end{figure}

\noindent\textbf{Impacts of different sizes of LLMs.} Next, we investigate the impacts of LLM sizes on the adaptation performance. We select OPT~\cite{zhang2022opt} as the foundational model for this investigation, which offers different versions with varying parameter sizes. The results are presented in Figure~\ref{fig:llm_sizes}. As shown, when the parameter size exceeds 1B, the adapted OPT achieves superior or comparable performance to the advanced learning-based algorithms. However, for the ABR task, OPT-0.35B performs significantly worse than all baselines, potentially due to the limited common knowledge to generalize across tasks. 
This suggests that, in practice, LLMs with parameter sizes greater than 1B are suitable for networking adaptation, while those smaller than 1B may not be the optimal choices for adaptation.

\noindent\textbf{Computation overhead.}  To measure the overhead of deploying \texttt{NetLLM}-adapted LLM to solve networking tasks, we profile the LLM answer generation process. 
Overall, loading a 7B LLM like Llama2-7B requires 29 GB memory and takes about 0.1s$\sim$0.3s to generate one answer. 
The computation overhead can be reduced by utilizing the advanced model compression techniques~\cite{deng2020model,xu2023survey} (discussed in $\S$\ref{sec:discussion}), or employing smaller LLMs such as OPT-1.3B which also achieves superior performance over baselines (see Figure~\ref{fig:llm_sizes}). 
Specifically, for OPT-1.3B, it only takes 7GB to load the model, which can be accommodated by commercial GPUs like NVIDIA 10GB 3080. Besides, it takes about 0.04s for OPT-1.3B to generate one answer, which is acceptable for many networking tasks.

\section{Discussion }
\label{sec:discussion}
\Revision{\textit{\textbf{Q1: What considerations are needed when adapting LLMs for specific networking tasks using \texttt{NetLLM}?}}}

\Revision{ While \texttt{NetLLM}'s overall design is independent of specific networking tasks, some considerations are needed when applying it for LLM adaptation. Specifically, when adapting LLMs for the target networking task, the creation of a new networking head is necessary, and the selection of a modality-specific feature encoder is also needed when dealing with a new modality. However, \texttt{NetLLM} minimizes the ad-hoc design costs associated with these considerations. On one hand, the networking head in \texttt{NetLLM} is essentially a simple linear layer that can be easily customized based on the task output space. For example, in VP task, the networking head is simply designed with three neurons to output the viewport coordinates, i.e., roll, pitch and yaw  (see $\S$\ref{appendix:netllm_impl}). On the other hand, \texttt{NetLLM} facilitates efficiency by allowing the reuse of existing encoders to process specific modalities, eliminating the need of handcrafting encoders from scratch. Furthermore, as the current LLM research landscape actively explores multimodality~\cite{openai2023gpt4,team2023gemini,fu2023mme,li2023blip}, \texttt{NetLLM} stands to benefit from the advancements in this field. With the development of more generic and powerful LLMs that support more modalities, \texttt{NetLLM} can utilize their built-in encoders to effectively process multimodal data.}

\vspace{0.2cm}
\noindent \Revision{\textit{\textbf{Q2: How does \texttt{NetLLM} compare to retrieval-augmented generation (RAG)?}}}

\Revision{Retrieval-augmented generation (RAG)~\cite{pan2024unifying,lewis2020retrieval,wu2022efficient} has been recently proposed to enhance the capabilities of LLMs. RAG constructs a large corpus as an external knowledge base to store domain-specific knowledge. During the inference stage, relevant information is retrieved from the knowledge base and attended to the input context of LLMs to provide domain knowledge. Although RAG has shown efficacy in improving the performance of LLMs in NLP tasks, it faces challenges when applied in the networking field. This is because, unlike NLP where knowledge can be stored explicitly in the textual form (e.g., the names of presidents of different countries), representing domain knowledge in networking as plain texts is challenging due to the abstract and implicit nature of networking knowledge (e.g., the ABR policy to dynamically adjust bitrates based on the changing network conditions). As a result, constructing an external knowledge base for RAG is rather challenging in the field of networking. In contrast, \texttt{NetLLM} takes a different approach by designing an efficient DD-LRNA module, which enables LLMs to automatically and effectively learn domain-specific knowledge for networking. With this module, \texttt{NetLLM} facilitates efficient acquisition of networking knowledge without relying on the construction of an external knowledge base.}

\vspace{0.2cm}
\noindent \Revision{\textit{\textbf{Q3: How to reduce the computation overhead of LLMs?}}}

\Revision{There is a wealth of research in model compression~\cite{deng2020model,xu2023survey} which can be leveraged  to reduce the computation overhead of LLMs, including model pruning~\cite{zhang2022advancing,liu2022discriminator}, quantization~\cite{tang2022arbitrary,tang2022mixed} and knowledge distillation~\cite{park2019rational,mirzadeh2022improved}. For example, 
SparseGPT~\cite{frantar2023sparsegpt} demonstrates that LLMs can be pruned to ignore at least 50\% parameters with minimal performance loss. 
OPTQ~\cite{frantar2022optq} uses quantization to decrease the bit-width of OPT-1.3B from originally 16 bits to 4 bits with negligible performance degradation while significantly reducing the model size by 4$\times$.
These active lines of research can be integrated into \texttt{NetLLM} to reduce the overhead of LLMs when deploying them for networking in practice.} \SecRevision{While the trade-off between performance and resource consumption should be considered when applying these techniques, we leave further exploration of this trade-off for future work.}

\vspace{0.2cm}
\noindent \SecRevision{\textit{\textbf{Q4: Why can LLMs be useful in networking?}}}

\SecRevision{In $\S$\ref{subsec:deep_dive}, we have identified from the high-level perspective that the pre-trained knowledge of LLMs is one of the dominant factors to their success in networking. However, further investigations into the internal working mechanisms of LLMs are crucial to improve their explainability. Gaining deeper insights into the explainability of LLMs enables researchers to comprehend their capabilities, limitations, and areas for improvement~\cite{zhao2024explainability,alammar2021ecco}. This understanding, in turn, paves the way for the development of more reliable and secure LLM-based networking systems that can be trustfully deployed in real-world scenarios. Therefore, a significant future research direction lies in designing an interpretable system to elucidate the behaviors of LLMs in the context of networking, which will greatly facilitate the effective utilization of LLMs in networking. }

\section{Concluding Remarks}
In this paper, we for the first time explore the utilization of LLMs as foundation models for networking to reduce handcraft costs involved in algorithm design and achieve strong generalization. 
To achieve this, we propose \texttt{NetLLM}, the first framework that efficiently adapts LLMs for different networking tasks. Across three use cases in networking, we show that \texttt{NetLLM}  enables the effective utilization of a single LLM to achieve superior performance and generalization in multiple networking tasks. While \texttt{NetLLM} by no means is the final answer, we hope that it serves as a stepping stone towards a more sustainable design philosophy for future networking algorithms and demonstrates the potential of adapting LLMs for networking.

\vspace{0.1cm}
\noindent\textbf{Ethics:} This work does not raise any ethical issues.

\section*{ACKNOWLEDGEMENTS}
We thank the anonymous SIGCOMM reviewers and our shepherd, Shay Vargaftik, for their invaluable feedbacks. This work was supported in part by NSFC with Grant No. 62293482, the Basic Research Project No. HZQB-KCZYZ-2021067 of Hetao Shenzhen-HK S\&T Cooperation Zone, NSFC with Grant No. 62102342, the Shenzhen Science and Technology Program with Grant No. RCBS20221008093120 047, the Young Elite Scientists Sponsorship Program of CAST with Grant No. 2022QNRC001, the Shenzhen Outstanding Talents Training Fund 202002, the Guangdong Research Projects No. 2017ZT07X1 52 and No. 2019CX01X104, the Guangdong Provincial Key Laboratory of Future Networks of Intelligence (Grant No. 2022B1212010001), and the Shenzhen Key Laboratory of Big Data and Artificial Intelligence (Grant No. ZDSYS201707251409055). Zhi Wang's work was  supported in part by Shenzhen Science and Technology Program (Grant No. JCYJ20220818101014030). Junchen Jiang's work was supported in part by NSF CNS 2146496, 1901466, 2313190, 2131826.

\newpage
\bibliographystyle{ACM-Reference-Format}
\bibliography{reference}
\appendix
\section{Appendices} 
Appendices are supporting material that has not been peer-reviewed.

\subsection{Details  of Figure~\ref{fig:challenge1and2}}
\label{appendix:measurement_setup}

\begin{figure}[t]
	\centering
	\includegraphics[width=0.45\textwidth]{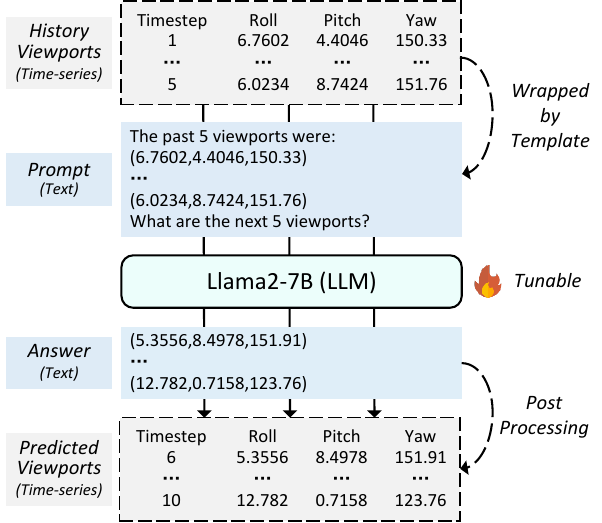}
	\vspace{-0.2cm}
	\caption{Illustration of using prompt learning~\cite{liu2023pre} to adapt the Llama2-7B LLM~\cite{touvron2023llama} for the VP task. 
 }
	\label{fig:prompt_and_finetune}
	\vspace{-0.2cm}
\end{figure}

The details to produce the results of ``Prompt Learning'' and ``Token Prediction'' in Figure~\ref{fig:challenge1and2} are illustrated in Figure~\ref{fig:prompt_and_finetune} and described as follows. We use prompt learning~\cite{liu2023pre} to adapt the Llama2-7B LLM for the VP task. Specifically, we design a \textit{prompt template} to encapsulate time-series viewports into textual prompts, and instruct Llama2~\cite{touvron2023llama} to generate answers for the VP task based on \textit{token prediction}. We conduct our measurements on an existing immersive video viewport dataset~\cite{jin2022where}. As a motivating example, we use Llama2 to predict future viewports in the next 1 second based on past viewports in the last 1 second. Following previous works~\cite{guimard2022deep,rondon2021track}, the viewport sampling rate is set to 5Hz, thus the future viewports and historical viewports are both 5-sample long. Note that Llama2 initially achieves poor performance without any fine-tuning. Hence, we further use OpenPrompt~\cite{ding2022openprompt}, an open-source prompt learning framework, to fine-tune Llama2 for 100000 iterations over the viewport dataset. 

Note that in Figure~\ref{fig:challenge1and2} (\textit{middle}), we calculate the fraction of valid answers generated by token prediction. In our practical implementation, we consider an answer as valid if we can extract viewports from it following a series of pre-defined string parsing operations. On the other hand, an invalid answer is often the case where it contains invalid characters (e.g., unexpected punctuation) or misses some values. While designing complex rules to post-process answers can alleviate the issue of invalid answers, it introduces the overhead of answer engineering~\cite{li2023chattwin}.

\subsection{Details of \texttt{NetLLM }Implementation}
\label{appendix:netllm_impl}
We have integrated \texttt{NetLLM} into three existing codebases for VP~\cite{wu2023mansy}, ABR~\cite{xia2022genet}, and CJS~\cite{decima_pytorch}. We have implemented the APIs in Figure~\ref{fig:api} based on the functionalities provided in the codebases. Additional details of \texttt{NetLLM} implementation are explained as follows.

For the multimodal encoder, we utilize ViT~\cite{dosovitskiy2021image} to encode images, and 1D-CNN~\cite{mao2017neural} to encode time-series and sequence data (e.g., historical throughputs and future chunk sizes at different bitrates in ABR). We leverage fully connected layer to extract features from scalar data (e.g., buffer occupancy in ABR), and use GNN~\cite{wu2021comprehensive,mao2019learning} to process graph information (e.g., DAGs in CJS). By default, the multimodal encoders are trainable, except that the parameters of ViT are frozen. This is because ViT has open source pre-trained weights which can be used to effectively extract image features.

The networking heads can be easily customized according to the target networking tasks. Specifically, we design the VP head to predict the viewport coordinates of roll, pitch, and yaw values. The ABR head is designed to output the probability distribution of candidate bitrates. As for CJS task, we design two heads for action generation: one to determine the next job stage to run and the other to decide the number of executor resources allocated to that stage. 

\SecRevision{Regarding the DD-LRNA scheme, we configure the context window $w$ for learning return distribution as 10 and 20 for ABR and CJS, respectively. We then set the rank $r$ of low-rank matrices to be 32, 128 and 128, for VP, ABR and CJS, respectively. }
While additional tuning of $w, r$ may be beneficial, we empirically find that \texttt{NetLLM} performs well across a wide range of hyperparameter values (generally, $w\ge 10$ and $r\ge32$ will yield good performance). Thus, we do not employ sophisticated methods to tune these hyperparameters and keep them fixed throughout the experiments in $\S$\ref{sec:evaluation}. \SecRevision{As for experience collection, we use GENET~\cite{xia2022genet} and Decima~\cite{mao2019learning} to collect experience datasets for the RL-based ABR and CJS tasks, respectively. While using more algorithms to interact with the environments for more epochs to expend the datasets may yield potential benefits,  we leave this for future exploration.}

\subsection{Overview of Baselines}
\label{appendix:baseline_overview}
In our evaluation, we compare the performance of the LLM adapted by our \texttt{NetLLM} framework with three baselines for each task, including state-of-the-art learning-based algorithms and rule-based algorithms. The following provides an overview of each baseline used in our evaluation. 

\begin{table}[t]
    \centering
    \caption{Summary of setting information in VP simulation. \textit{hw}/\textit{pw} is short for historical window/prediction window.}
    \label{table:vp_settings}
    \vspace{-0.3cm}
    \begin{threeparttable}
    \small
    \begin{tabular}{m{2cm}<{\centering}  m{2.8cm}<{\centering}  m{2.8cm}<{\centering}  }
    \toprule
         \textbf{Setting} & \textbf{Viewport Dataset} & \textbf{Prediction Setup}  \\
         \midrule
         \textit{default train} & \textit{Jin2022} & \textit{hw} =2s, \textit{pw} = 4s \\
         \textit{default test} & \textit{Jin2022} & \textit{hw} = 2s, \textit{pw} = 4s \\
         \textit{unseen setting1} & \textit{Jin2022} & \textit{hw} = 4s, \textit{pw} = 6s \\
         \textit{unseen setting2} & \textit{Wu2017} & \textit{hw} = 2s, \textit{pw} = 4s \\
         \textit{unseen setting3} & \textit{Wu2017} & \textit{hw} = 4s, \textit{pw} = 6s \\
         \bottomrule
    \end{tabular}
    \end{threeparttable}
    \vspace{-0.3cm}
\end{table}

\noindent\textbf{Baselines for VP.} We implement the following three baselines for performance comparison for the VP task: TRACK~\cite{rondon2021track}, linear regression (labeled "LR")~\cite{qian2018flare}, and velocity-based prediction (labeled "Velocity")~\cite{feng2021liveobj}. TRACK~\cite{rondon2021track} is a learning-based algorithm that designs a DNN model based on Long Short Term Memory (LSTM) architecture for VP. It considers both viewer's historical viewports and video saliency map as inputs to achieve state-of-the-art performance, where saliency map is an image that describes viewer's potential attention on the video content. LR~\cite{qian2018flare} assumes the movement of viewer's viewports as a linear function related to time, then uses linear regression to estimate such function for predicting viewer's viewports. Velocity~\cite{feng2021liveobj} calculates the moving speed of viewer's historical viewports and uses it to estimate the positions of viewer's future viewports.

\SecRevision{Note that since the open-source codes of TRACK are originally written in Keras~\cite{keras}, we carefully convert its codes into PyTorch~\cite{pytorch} to make TRACK compatible with the VP codebase~\cite{wu2023mansy}. We have ensured that our implementation preserves the same functionality of TRACK as its original implementation. Besides, as TRACK does not offer pre-trained model weights, we re-train it from scratch on our datasets with the same training hyperparameters described in TRACK's paper and codes. As for rule-based algorithms LR and Velocity, which do not provide open source implementation, we implement them ourselves  by strictly following the same ideas and formulas presented in their respective papers.}

\noindent\textbf{Baselines for ABR.} As for ABR, the following baselines are implemented for comparison: GENET~\cite{xia2022genet}, BBA~\cite{huang2014buffer} and MPC~\cite{yin2015control}. GENET~\cite{xia2022genet} is a RL-based streaming algorithm improved over Pensieve~\cite{mao2017neural}. It introduces a curriculum learning technique to facilitate the RL training process to improve convergence performance. BBA~\cite{huang2014buffer} considers buffer occupancy as a critical signal for bitrate control and designs an algorithm to maintain the playback buffer occupancy at a desired level. MPC~\cite{yin2015control} leverages both throughput estimates and buffer occupancy to choose bitrates by optimizing a given QoE metric over a future chunk horizon. 

\SecRevision{To implement the aforementioned ABR baselines, we utilize the open-source codes of GENET~\cite{xia2022genet}, which already include the implementation of the three algorithms. Furthermore, we re-use the pre-trained model weights of GENET\footnote{\url{https://github.com/GenetProject/Genet/tree/main/src/emulator/abr/pensieve/data/mahimahi_new_best_models/ADR_model}} for our experiments.}

\begin{table}[t]
    \centering
    \caption{Summary of setting information in ABR simulation.}
    \label{table:abr_settings}
    \vspace{-0.3cm}
    \begin{threeparttable}
    \small
    \begin{tabular}{m{2cm}<{\centering}  m{2.8cm}<{\centering}  m{2.8cm}<{\centering}  }
    \toprule
         \textbf{Setting} & \textbf{Video Dataset} & \textbf{Bandwidth Traces} \\
         \midrule
         \textit{default train} & \textit{Envivio-Dash3} & \textit{FCC}\\
         \textit{default test} & \textit{Envivio-Dash3} & \textit{FCC} \\
         \textit{unseen setting1} & \textit{Envivio-Dash3} & \textit{SynthTrace} \\
         \textit{unseen setting2} & \textit{SynthVideo} & \textit{FCC} \\
         \textit{unseen setting3} & \textit{SynthVideo} & \textit{SynthTrace}\\
         \bottomrule
    \end{tabular}
    \end{threeparttable}
    \vspace{-0.3cm}
\end{table}

\noindent\textbf{Baselines for CJS.} The following three baselines are implemented for  the CJS task: Decima~\cite{mao2019learning}, first-in-first-out scheduling (labeled "FIFO")~\cite{spark_scheduler} and fair scheduling (labeled "Fair")~\cite{spark_scheduler}. Decima~\cite{mao2019learning} is a RL model for job scheduling in the distributed computing cluster, which develops a graph neural network (GNN) to efficiently process DAG information of job properties (e.g., resource demands and dependency). Both FIFO and Fair are two common scheduling algorithms used by data processing system Spark~\cite{spark_scheduler}. The former schedules jobs in the order of their arrival and allocates the requested amount of resources to each job, while the latter schedules jobs in a ``round robin'' fashion to ensure that each job receives a roughly equal share of the cluster.

\SecRevision{We utilize the PyTorch re-implementation of Decima~\cite{decima_pytorch} for our experiments as the original implementation is somewhat outdated. Furthermore, we make use of the pre-trained model weights of Decima\footnote{\url{https://github.com/ArchieGertsman/spark-sched-sim/tree/main/models/decima}} provided in~\cite{decima_pytorch}. Additionally, we adopt the implementation of FIFO and Fair from the same source~\cite{decima_pytorch}.}

\subsection{Details of Simulation Settings}
\label{appendix:simulation_settings}

\begin{table}[t]
    \centering
    \caption{Summary of setting information in CJS simulation.}
    \label{table:cjs_settings}
    \vspace{-0.3cm}
    \begin{threeparttable}
    \small
    \begin{tabular}{m{2cm}<{\centering}  m{2.8cm}<{\centering}  m{2.8cm}<{\centering} }
    \toprule
         \textbf{setting} & \textbf{Job Requests} & \textbf{Executor Resources(k)} \\
         \midrule
         \textit{default train} & 200 & 50 \\
         \textit{default test} & 200 & 50 \\
         \textit{unseen setting1} & 200 & 30 \\
         \textit{unseen setting2} & 450 & 50 \\
         \textit{unseen setting3} & 450 & 30 \\
         \bottomrule
    \end{tabular}
    \end{threeparttable}
    \vspace{-0.3cm}
\end{table}

We generate different simulation environments with real-world and synthetic datasets for training and testing to comprehensively evaluate the performance of the LLM adapted by \texttt{NetLLM} against baselines. The  detailed simulation settings for each task are explained as follows.


\noindent\textbf{VP simulation.} As shown in Table~\ref{table:vp_settings}, by default, we train and test each method on a large-scale viewport dataset \textit{Jin2022}~\cite{jin2022where} which records the viewport traces from 84 viewers\footnote{\SecRevision{The \textit{Jin2022} dataset originally contains the viewport traces from 100 viewers~\cite{jin2022where}. We filter out those incomplete ones (i.e., less than 60 seconds in duration) and finally use the traces from 84 viewers for experiments.}} watching 27 60-second immersive videos. \SecRevision{We randomly select 15 videos and 42 viewers for training, 6 videos and 21 viewers for validation, 6 videos and 21 viewers for testing. This  results in a total of 882 traces for experiments.} The historical window (\textit{hw}) and prediction window (\textit{pw}) are set to be 2 seconds and 4 seconds, respectively, for the default training and testing settings.

When evaluating generalization performance, we test each method on a new viewport dataset (i.e., new data distributions) and/or with a new prediction setup (i.e., increasing prediction difficulty). For instance, on \textit{unseen setting2}, we evaluate each method on the new \textit{Wu2017} dataset~\cite{wu2017dataset}. This dataset contains 9 videos\footnote{\SecRevision{The \textit{Wu2017} dataset originally includes 18 videos. We use the first 9 videos as viewers are free to look around when watching these videos.}} with an average length of 242 seconds watched by 48 viewers. \SecRevision{We randomly sample 4 videos and 9 viewers from the dataset, resulting in 36 long viewport traces  for testing generalization.} As for \textit{unseen setting1}, we increase \textit{pw} to increase the prediction difficulty for each method. Following the setting in~\cite{qian2018flare}, we roughly set \textit{hw} $ \approx$ \textit{pw} / 2 across all settings. Changing the coefficient does not qualitatively affect the results.

\noindent\textbf{ABR simulation.} Table~\ref{table:abr_settings} summarizes the simulation settings for ABR. By default, we train and test all methods to stream the \textit{Envivio-Dash3} video from the DASH-246 JavaScript reference client~\cite{dashvideo}, whose format follows the GENET~\cite{xia2022genet} and Pensieve~\cite{mao2017neural} setting. We use the broadband \textit{FCC}~\cite{fcc} traces as the default bandwidth dataset. 
\SecRevision{In particular, we use the same traces for training and validation as those used by GENET, which compromise 235 traces for training and 150 traces for validation. Then, we randomly sample 100 traces from the remaining dataset for testing. This results in the use of more than 90 hours of bandwidth traces for experiments.} 
To simulate environments for generalization testing, we follow the method in Pensieve~\cite{mao2017neural} to generate a synthetic video \textit{SynthVideo} which shares a similar format of \textit{Envivio-Dash3} but with a larger video bitrate. Besides, we also generate a new bandwidth dataset \textit{SynthTrace} \SecRevision{with 100 traces} according to the method in Pensieve~\cite{mao2017neural}, which exhibits a larger bandwidth range and more dynamic fluctuation patterns than \textit{FCC}.

\noindent\textbf{CJS simulation.} Table~\ref{table:cjs_settings} provides the detailed information of the CJS simulation. Following Decima~\cite{mao2019learning}, we simulate different workload traces using a real-world dataset TPC-H~\cite{chiba2016workload} which contains job requests of large data volumes, high executor demands, and high degree of complexity. 
\SecRevision{To be consistent with the settings used by the pre-trained Decima in~\cite{decima_pytorch}}, we set the number of job requests to be 200 and the number of executor resources (representing computation resources) to be 50k units as the default training and testing settings.
To evaluate the generalization performance of each method, we simulate various unseen harder workloads by increasing the number of job requests and reducing the number of executor resources, as also done in Decima~\cite{mao2019learning}. \SecRevision{Note that in each setting the job requests are randomly sampled from the TPC-H dataset. Besides, when evaluating on the default testing setting, we have ensured that the job requests are different from those in the training setting. This can be easily done by setting different random seeds for data sampling.}


\subsection{Real-world ABR Testbed Setup}
\label{appendix:real_world}
We leverage the testbed from GENET~\cite{xia2022genet} to test the \texttt{NetLLM}-adapted Llama2 in a real-world client-server ABR system. The testbed modifies dash.js (version 2.4) to support BBA, MPC and GENET streaming algorithms. We further modify the dash.js to support the adapted Llama2. In our real-world tests, the client video player is a Google Chrome browser (version 87) and the video server (Apache version 2.7) runs on the same machine as the client. All tests are performed on our Linux server, with two different ports to emulate the ABR client and video server. We then use Mahimahi~\cite{netravali2015mahimahi} to emulate different network environments from the broadband traces~\cite{fcc} and cellular mobile traces~\cite{norway}, along with an 80ms RTT, between the client and server. \SecRevision{In particular, we randomly sample 100 traces from both the broadband and cellular mobile bandwidth datasets for network environment emulation.}

\subsection{Evaluation Metrics}
\label{appendix:eval_metrics}
\noindent\SecRevision{\textbf{Metric for VP.} We use mean absolute error (MAE) as the evaluation metric for the VP task. Let $\boldsymbol{v}_p=(\alpha_t, \beta_t, \zeta_t)$ denote the viewport coordinate at timestep $t$, where $\alpha, \beta, \zeta$ represent the roll, pitch and yaw values, respectively. Given $\boldsymbol{v}^p_t, \boldsymbol{v}^g_t$ as the predicted and ground-truth viewports and $H$ as the prediction horizon, the MAE is calculated by:
$$MAE=\frac{1}{H}\sum_{t=1}^H \frac{|\alpha_t^p-\alpha_t^g| + |\beta_t^p-\beta_t^g| + |\zeta_t^p-\zeta_t^g)|}{3}$$}

\noindent\SecRevision{\textbf{Metric for ABR.} We use quality of experience (QoE) as the evaluation metric for the ABR task, which is defined as the weighted linear combination of three metrics~\cite{mao2017neural,xia2022genet}:
$$QoE=\frac{\sum_{i=1}^C(Bitrate_i - \lambda Rebuf_i - \gamma BitrateChange_i)}{C}$$
where $Bitrate_i$ is the bitrate in Mbps of chunk $i$, $Rebuf_i$ is the rebuffering time in seconds of downloading chunk $i$, $BitrateChange_i$ is the bitrate change in Mbps between consequtive chunks, $C$ is the number of chunks of the video and $\lambda, \gamma$ are the weight parameters. Following Pensieve~\cite{mao2017neural}, we set  $\lambda=4.3$ and $\gamma = 1$.}

\noindent\SecRevision{\textbf{Metric for CJS.} We use job completion time (JCT)~\cite{mao2019learning} as the evaluation metric for the CJS task. Let $t_s$ denote the arrival time of a job and $t_e$ denote the finishing time of a job. The JCT is calculated by:
$$JCT=t_e-t_s$$}

\end{document}